\documentstyle[11pt,axodraw]{article}

\newcommand{\ds}{\displaystyle}

\newcommand{\nn}{\nonumber}
\newcommand{\cO}{{\cal O}}
\newcommand{\cL}{{\cal L}}
\newcommand{\cQ}{{\cal Q}}
\newcommand{\cA}{{\cal A}}
\newcommand{\cP}{{\cal P}}
\newcommand{\cC}{{\cal C}}
\newcommand{\cT}{{\cal T}}
\newcommand{\cH}{{\cal H}}
\newcommand{\cM}{{\cal M}}

\def\bc{\begin{center}}
\def\ec{\end{center}}
\def\be{\begin{equation}}
\def\ee{\end{equation}}
\def\bea{\begin{eqnarray}}
\def\eea{\end{eqnarray}}
\def\ra{\rightarrow}

\def\dg{\dagger}
\def\kmung{$K~\rightarrow~\mu\nu\gamma$}
\def\GF{G_{\mbox{\tiny F}}}

\def\xio{\xi_{\mbox{\tiny odd}}}
\def\xiem{\xi_{\mbox{\tiny EM}}}
\def\im{\mbox{\tt I\hspace*{-2pt}m}}
\def\re{\mbox{\tt R\hspace*{-2pt}e}}
\def\CP{$\cC\!\cP$}

\def\bdm{\begin{displaymath}}
\def\ea{\end{array}}
\def\edm{\end{displaymath}}
\def\ds{\displaystyle}

\font\cha=cmitt10 scaled\magstep2
\font\chp=cmtt10 scaled\magstep1

\def\etazz{{\langle \pi^0\pi^0 | K^0_L \rangle \over \langle \pi^0\pi^0 | K^0_S \rangle}}
\def\etapm{{\langle \pi^+\pi^- | K^0_L \rangle \over \langle \pi^+\pi^- | K^0_S \rangle}}

\begin{document}
\bc
{\huge \bf \CP \  Violation in $K$-Meson Decays}\\[3mm]

{\large R. N. Rogalyov}\footnote{E-mail: rogalyov@mx.ihep.su}  \\[2mm]
State Research Center of Russia\\
"Institute for High-Energy Physics", Protvino, Russia  \\[4mm]
\ec

\begin{abstract}
Here we present a pedagogical review of \CP\ and $\cT$ violation
in the decays of $K$ mesons. Diagonalization of the quark mass
matrix and the emergence of the complex phase in both the standard
and the left--right symmetric models is considered in great detail.
A special emphasis is focused on a correct definition of \CP-violating
quantities: $\epsilon, \epsilon'$ {\it etc.} 
(with due regard for the Wu--Yang phase convention) and
to formulation of the time-reversal invariance criterion in the
elementary particle physics.
A particular attention has been concentrated on theoretical
evaluation of the parameters $\epsilon$ and $\epsilon'$ and
the \CP- and $\cT$-violating asymmetries in the decays $K\ra\mu\nu\gamma$
and $K\ra 3\pi$.
\end{abstract}

\section{CKM Matrix}

\subsection{Diagonalization of the Mass Matrix.}

The initial Lagrangian of the Standard Model (SM) involves 12 massless chiral quarks

\hspace*{9mm}\begin{tabular}{cc}
& \\
 $D^R_1$ & $U^R_1$ \\ 
& \\ 		& \\
\ 
 $D^L_1$ &  $U^L_1$  \\ \hline
& \\	
& \\
 $D^R_2$ & $U^R_2$ \\ 
& \\ 		& \\
  $D^L_2$ & $U^L_2$  \\ \hline
& \\	
& \\
 $D^R_3$ & $U^R_3$ \\ 
& \\ 		& \\
 $D^L_3$ &  $U^L_3$  \\ \hline
& \\	
\end{tabular}
\hspace*{19mm} ${\bf \sim}$ \hspace*{19mm} \begin{tabular}{cc}
& \\
 $d^R$ & $u^R$ \\ 
& \\ 		& \\
\ 
 $d^L$ & $u^L$  \\ \hline
& \\	
& \\
 $s^R$ & $c^R$ \\ 
& \\ 		& \\
\ 
 $s^L$ & $c^L$  \\ \hline
& \\	
& \\
 $b^R$ & $t^R$ \\ 
& \\ 		& \\
\ 
 $b^L$ & $t^L$  \\ \hline
& \\	
\end{tabular}

\noindent which acquire mass due to spontaneous breaking of the $SU(2)_L$ symmetry.
The most general quark mass matrix induced by the Higgs fields has the form
\be\label{InitSMMassTerm}
\sum_{i,j=1}^3 
\left( \bar D_i^L M'_{ij} D_j^R +  \bar U_i^L N'_{ij} U_j^R  \right) + \mbox{H.c.},
\ee
where $M'_{ij}$ and $N'_{ij}$ are arbitrary $3\times 3$ matrices.
This mass term can be considered as a perturbation of the initial
SM Hamiltonian, the above quark states form the basis associated with 
the 12-fold degenerate zero-mass level of the unperturbed Hamiltonian. 
According to the quantum mechanics,
one should find the basis in which the perturbation operator takes the diagonal 
form
\be
M'=V_dMY_d^\dagger, \ \ \ \ \ N'=V_uMY_u^\dagger, 
\ee
where $M=\mbox{diag}(m_d, m_s, m_b)$ and $N=\mbox{diag}(m_u, m_c, m_t)$;
$Y_d$, $Y_u$, $V_d$, and $V_u$ are unitary $3\times 3$ matrices; and
\be
d^R_i, \ u^R_i, \ d^L_i,\ u^L_i, \ (i=1,2,3)
\ee
are the so called mass eigenstates, which form the sought-for basis.
The interaction eigenstates (the eigenstates of the interaction Hamiltonian)
are expressed in terms of the mass eigenstates as follows:
\bdm \begin{array}{cc}
D^R=Y_d\;d^R, \ \ \ & \ U^R=Y_u\;u^R, \\ 
D^L=V_d\;d^L, \ \ \ & \ U^L=V_u\;u^L. 
\ea \edm
This being so, the interaction between left charged currents and gauge
bosons
\be
\sum_{i=1}^3 
\left( \bar D_i^L \hat W^- U_i^L +  \bar U_i^L \hat W^+ D_i^L  \right)
\ee
takes the form
\be
 \sum_{i,j=1}^3 
 \bar d_i^L (V_d^\dagger V_u)_{ik} \hat W^- u_j^L +  
 \bar u_i^L (V_u^\dagger V_d)_{ik} \hat W^+ d_j^L \ = \  
 \sum_{i=1}^3 \bar u_i \gamma^\mu {1-\gamma^5 \over 2}W^+_\mu d'_i + \mbox{H.c.},
\ee
where 
\be
d'_i=V_{ij} d_j, \ \ \mbox{and} \ \ \ V=V_u^\dagger V_d
\ee
is the Cabibbo--Kobayashi--Maskawa (CKM) matrix \cite{ChengLi}. {\bf Note that the
matrices $Y_d$ and $Y_u$ play no role when considering interaction.}
Now the left doublets (that appear in the interaction Lagrangian)
have the form
\bdm
\left(\begin{array}{c}
u	\\
V_{ud}d+V_{us}s+V_{ub}b
\ea\right), \ \ \
\left(\begin{array}{c}
c	\\
V_{cd}d+V_{cs}s+V_{cb}b
\ea\right), \ \ \ 
\left(\begin{array}{c}
t	\\
V_{td}d+V_{ts}s+V_{tb}b
\ea\right).
\edm
It should be noticed that 
mass eigenstates are invariant under the transformations
$V_u \ra V_u\Phi_u$, $V_d \ra V_d\Phi_d$, where
\bdm
\Phi_u=\left(
\begin{array}{ccc}
e^{i\phi_u} & 0 & 0 \\ 
0 & e^{i\phi_c} & 0 \\ 
0 & 0 & e^{i\phi_t}  
\ea  \right) ,\ \ \ \ \ 
\Phi_d=\left(
\begin{array}{ccc}
e^{i\phi_d} & 0 & 0 \\ 
0 & e^{i\phi_s} & 0 \\ 
0 & 0 & e^{i\phi_b}  
 \ea \right);
\edm
thus any matrix of the family $\Phi_u^\dagger V \Phi_d$ may be chosen as the 
quark-mixing matrix. To exclude this arbitrariness, one should fix the phases
of the quark states:
\bdm
\mbox{fixing the phase of the} \hspace*{10mm}
\begin{array}{c}
u\\ 
c\\ 
t\\ 
s\\ 
b\\ 
\ea
\hspace*{10mm} \mbox{quark \ \ \ \ allows to make } \hspace*{12mm} 
\begin{array}{c}
V_{ud} \\ 
V_{cd} \\ 
V_{td} \\ 
V_{us} \\ 
V_{ub} \\ 
\ea
\hspace*{12mm} \mbox{real.}\\
\edm
The phase of the $d$ quark is fixed by the requirement that the above matrix elements are real. 

Thus 5 independent conditions can be imposed on the elements of the mixing matrix 
and so it has 4 independent parameters, 1 of which has to be complex (all real-valued 
$U(3)$ matrices belong to the $SO(3)$ group). This means that there is one \CP - and 
$\cT$- violating parameter in the SM interaction Lagrangian, whose value cannot be
determined from the general principles. In the case of $N$ flavors, a similar reasoning
implies the existence of $(N-1)(N-2)/2$ complex phases; in the case of two generations, 
all the elements of the mixing matrix can be made real and the theory is \CP \ invariant.

It is instructive to show in detail how the imaginary part
of the fermion--fermion--vector-boson coupling constant
implies \CP\ and $\cT$ violation. We define the action of the $\cC$, $\cP$, and $\cT$
transformations on the fermion fields as follows:
\be
\cP\psi(x)\cP^\dagger = \gamma^0\psi(\tilde x),\hspace*{10mm} 
\cC\psi(x)\cC^\dagger = -i \gamma^2\psi^\ast(\tilde x),\hspace*{10mm}  
\cT\psi(x)\cT^\dagger = i\gamma^1\gamma^3\psi(-\tilde x),
\ee
where $\tilde x^\mu = x_\mu$. Then the action of these operators on the fermion densities
\bea
S(x)&=&:\bar q(x) q(x): \\ \nonumber
V^\mu(x)&=&:\bar q(x) \gamma^\mu q(x): \\ \nonumber
T^{\mu\nu}(x)&=&:\bar q(x) \sigma^{\mu\nu} q(x): \\ \nonumber
A^\mu(x)&=&: \bar q(x) \gamma^\mu\gamma^5 q(x): \\ \nonumber
P(x)&=&: i\bar q(x) \gamma^5 q(x):  \nonumber
\eea
takes the form  \cite{Zuber} (note that $\cT$ conjugates all complex constants):\\[3mm]

{\bf Table 1.} $\cP, \cC,$ and $\cT$ transformations of the fermion densities. \\[2mm]

\hspace*{-3mm}\begin{tabular}{|c|c|c|c|c|c|}
\hline
&&&&& \\
Transformation & $S(x)$ & $V^\mu(x)$ & $T^{\mu\nu}(x)$ & $ A^\mu(x)$ &  $P(x)$ \\
 &&{\small and vector field}&&& \\  \hline 

&&&&& \\
$\cP$ & $S(\tilde x)$ & $V_\mu(\tilde x)$ & $T_{\mu\nu}(\tilde x)$ & $ A_\mu(\tilde x)$ &  $-\; P(\tilde x)$ \\
&&&&& \\  \hline 

&&&&& \\
$\cC$ & $\ \ \ S(x)\ \ \ $ & $\ \ -\; V^\mu(x)\ \ $ & $\ \ -\; T^{\mu\nu}(x)\ \ $ & $ \ \ -\; A^\mu(x)\ \ $ &  $\ \ \ P(x)\ \ \ $ \\
&&&&& \\  \hline 

&&&&& \\
$\cT$ & $S(-\; \tilde x)$ & $V_\mu(-\; \tilde x)$ & $-\; T_{\mu\nu}(-\; \tilde x)$ & $ A_\mu(-\; \tilde x)$ &  $-\; P(-\; \tilde x)$ \\
&&&&& \\  \hline 

&&&&& \\
$\Theta =\cC \cP \cT$ & $S(-\; x)$ & $-\; V^\mu(-\; x)$ & $T^{-\; \mu\nu}(x)$ & $ -\; A^\mu(-\; x)$ &  $P(x)$ \\
&&&&& \\  \hline 
\end{tabular}
\vspace*{3mm}

These definitions agree with the transformations of various physical quantities
under $\cC$, $\cP$, and $\cT$ (see Table 4).
It is seen now that the term,\footnote{Operator $\bar t...d$ creates $t$ and anti-$d$ quarks and annihilates $d$ and anti-$t$ quarks.} say, 
\be\label{eq:CPoddTerm}
\int dx\ \left( V_{td}\;\bar t\gamma^\mu(1-\gamma^5) d\ W^+_\mu\ + \ V_{td}^\ast\;\bar d\gamma^\mu(1-\gamma^5) t \ W^-_\mu \right)
\ee
is both \CP\ and $\cT$ violating due to imaginary part of $V_{td}$: \CP\ transformation 
changes the fermion densities: $\bar t\gamma^\mu(1-\gamma^5) d\ W^+_\mu\ 
\leftrightarrow  \bar d\gamma^\mu(1-\gamma^5) t \ W^-_\mu $, whereas $\cT$ has no effect on the
operators, however, conjugates the coefficients: $V_{td}\; \leftrightarrow  V_{td}^\ast$.
Note that the $\cC \cP \cT$ transformation leaves the term (\ref{eq:CPoddTerm}) invariant.

\subsection{The strong \CP\  problem.}
Strictly speaking, one could fix the phases of left-handed and right-handed states separately,
excluding arbitrariness in the chiral phases $\alpha$:
\be
q^R \ra e^{i\alpha} q^R, \ \ \ \ q^L \ra e^{-i\alpha} q^L.
\ee
This would give rise to the terms 
\be\label{eq:PS}
i h^u_1\bar u\gamma^5 u + i h_1^d\bar d\gamma^5 d + ... 
\ee
in the interaction Lagrangian. 
A strong limit on the \CP-odd parameters $h_i$ comes from the mechanism
of the spontaneous breaking of the chiral $SU_L(N_f)\times SU_3(N_f)$ symmetry in 
strong interactions. Let us assume that the vacuum is \CP-even.
In this case, the v.e.v. of the mass term
\be
H_m'=\sum_{i=1}^{N_F} \bar q_i(m_i - ih_i \gamma^5)q_i
\ee
can approach its minimum only provided that $h_u=h_d=h_s=...=\lambda$ (a consequence of the
Dashen theorem \cite{Dashen}). If we fix the chiral phases of quarks so that the
mass term is $\gamma^5$-free, we have to consider the spontaneous 
breaking of the \CP\ symmetry:
\be
\langle 0|\bar U^L_i U^R_j |0 \rangle = \langle 0|\bar D^L_i D^R_j |0 \rangle = Ce^{i\phi_i}\delta_{ij},\ \ \
\ee
where the angles $\phi_i$ are $\cC\!\cP$-violating parameters. 
Thus we got rid of the $SU(N_f)$ chiral phases.
The $U_A(1)$ chiral phase can be 
absorbed in the so called $\theta$ term giving rise to the strong $\cC\!\cP$ violation:
the measurable quantities are independent of $\theta - 2N_F\alpha$,
where $\alpha$ is the chiral phase and $\theta$ is the coefficient of the 
\CP-odd term \cite{Crewther}
\be
\cL_\theta = {\theta \over 32\pi^2} G_{\mu\nu}G^{\mu\nu},
\ee
which is equivalent to (that is, can be replaced with) the "pseudoscalar mass term"
\be
\ds \sum_{i=1}^{N_F} \  {i\over\ds \left({1\over m_1} + ... + {1\over m_{N_F}}\right)} \ \ \bar q_i  \gamma^5 q_i.
\ee

\subsection{The CKM Matrix}

Performing the above procedure, we arrive at the expression for the quark 
mixing matrix in the form proposed by Kobayashi and Maskawa:
\bdm
\left(
\begin{array}{ccc}
V_{ud} & V_{us} & V_{ub} \\ 
V_{cd} & V_{cs} & V_{cb} \\ 
V_{td} & V_{ts} & V_{tb}  
\ea  \right) \ 
=\ \left(
\begin{array}{ccc}
c_1 & -s_1 c_3 & -s_1 s_3 \\ 
s_1 c_2 & c_1 c_2 c_3 - s_2 s_3 e^{i\delta} & c_1 c_2 s_3 + s_2 c_3 e^{i\delta} \\ 
s_1 s_2 & c_1 s_2 c_3 + c_2 s_3 e^{i\delta} & c_1 s_2 s_3 - c_2 c_3 e^{i\delta}  
 \ea \right),
\edm
where $c_i=\cos\theta_i, s_i=\sin\theta_i$.
This matrix differs from the mixing matrix advocated by the Particle Data Group \cite{PDG}, 
\bdm
\left(
\begin{array}{ccc}
c_{12}c_{13} & s_{12}c_{13} & s_{13}e^{-i\delta_{13}} \\ 
-s_{12}c_{23} - c_{12}s_{23}s_{13}e^{i\delta_{13}} & 
c_{12}c_{23} - s_{12}s_{23}s_{13}e^{i\delta_{13}} & s_{23}c_{13} \\ 
s_{12}s_{23} - c_{12}c_{23}s_{13}e^{i\delta_{13}} & 
-c_{12}s_{23} - s_{12}c_{23}s_{13}e^{i\delta_{13}} & c_{23}c_{13}
 \ea \right),
\edm
in the phases of the $c, b, t$ quarks. A popular approximation
that emphasizes the hierarchy in the size of the angles 
\be
s_{12}>\!\!>s_{23}>\!\!>s_{13},
\ee
where $s_{12}\equiv\lambda$ is the sine of the Cabibbo angle,
is that one expands the other elements in terms of the parameter $\lambda$.
Up to and including terms of order $\lambda^3$, the mixing matrix is given by
\bdm
\left(
\begin{array}{ccc}
1-{\lambda^2 \over 2} & \lambda & A\lambda^3(\rho-i\eta) \\ 
-\,\lambda & 1-{\lambda^2 \over 2} & A\lambda^2 \\ 
 A\lambda^3(1-\rho-i\eta) & -\;A\lambda^2  & 1
\ea  \right), \ 
\edm
where $A, \rho, \eta$ are assumed to be of order unity.
The unitarity of the CKM matrix implies 
\be
V_{ub}^\ast V_{ud} + V_{cb}^\ast V_{cd} + V_{tb}^\ast V_{td} = 0.
\ee
In the approximation $V_{ud}\simeq V_{tb} \simeq 1$ one obtains
\be
{V_{ub}^\ast \over A\lambda^3} + {V_{td}\over A\lambda^3} - 1 = 0.
\ee
This relation identifies a triangle in the $\rho -\eta$ plane (see Fig.~1); the angles of this triangle are
\bc
\begin{figure*}[hhh] \hbox{
\hspace*{-5pt}
       \epsfxsize=350pt \epsfbox{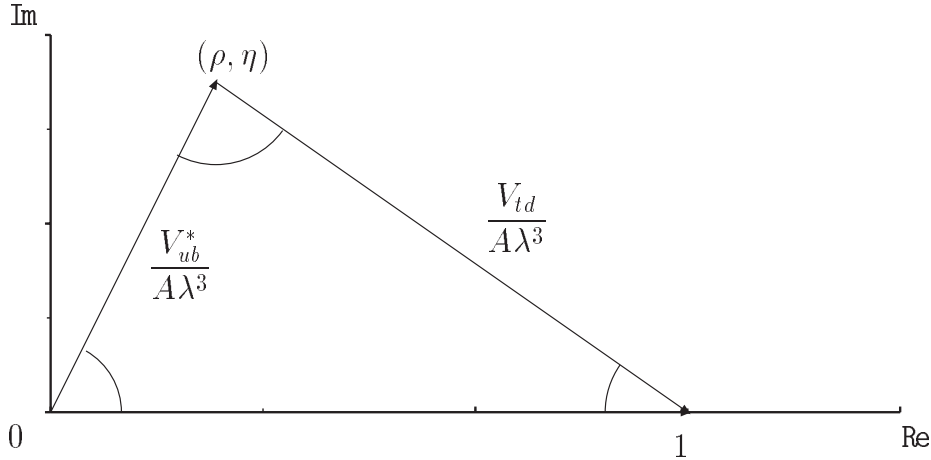} \hspace*{10pt}
       }
\caption{The unitarity triangle.} \label{fig:1}
\end{figure*}
\ec
measures of $\cC\!\cP$ violation. The origin of the complex CKM phase
is still not clearly understood. It may be that it stems from the
short-distance dynamics \cite{Volkov} or extra dimensions (or something else)
\cite{Mohapatra01}.

\subsection{Left-Right Symmetric Model}
The fermion sector of the $SU(3)_L\times SU(3)_R \times U(1)$-symmetric 
model \cite{Pati} is the same as that of the SM; the boson sector involves the left
$W^L_\mu$ and right $W^R_\mu$ gauge bosons, whose interaction with quarks
$\bar q \hat D q$ is determined by the covariant derivative
\be 
D_\mu q= \partial_\mu q + i g_L\vec W^L_\mu \vec \tau {1-\gamma^5 \over 2} q
		+ i g_R\vec W^R_\mu \vec \tau {1+\gamma^5 \over 2} q,
\ee
and a suitable $SU(3)_L\times SU(3)_R$ multiplet of the Higgs fields 
\bdm
\left(
\begin{array}{cc}
\phi_{11} & \phi_{12} \\
\phi_{21} & \phi_{22} 
\ea\right)
\edm
interacts with quarks as follows:
\be 
\bar q^L_i (h_{ij}\phi + f_{ij}\tilde\phi) q^R_j + \mbox{H.c.},
\ee
where $\tilde\phi = -\sigma_2\;\phi^\ast\;\sigma_2$ and the indices $i,j$
denote generation. Vacuum expectation values (v.e.v.) of the Higgs fields can be taken to be
\bdm
\langle \phi \rangle = 
\left(
\begin{array}{cc}
\kappa & 0 \\
0 & \kappa' 
\ea\right)
\edm
giving rise to the mass term of the form
\bdm
\sum_{i,j=1}^3
\left(\bar U^L_i, \bar D^L_i\right)
\left(
\begin{array}{cc}
h_{ij}\kappa + f_{ij}\kappa' & 0 \\
0 & h_{ij}\kappa' + f_{ij}\kappa
\ea\right)
\left( \begin{array}{c}
U_j^R \\
D_j^R
\ea \right)\ \ = \ \ 
\bar D^L_i M'_{ij} D^R_j + \bar U^L_i N'_{ij} U^R_j,
\edm
where $M'_{ij}=f_{ij}\kappa + h_{ij}\kappa'$ and $N'_{ij}=f_{ij}\kappa' + h_{ij}\kappa$.
Diagonalization of the mass matrices $M'$ and $N'$ can be made by the same token
as in the case of the SM: 
\bdm \begin{array}{cc}
D^R=Y_d\;d^R, \ \ \ & \ U^R=Y_u\;u^R, \\ 
D^L=V_d\;d^L, \ \ \ & \ U^L=V_u\;u^L,
\ea \edm
where
\be
d^R_i, \ u^R_i, \ d^L_i,\ u^L_i, \ (i=1,2,3)
\ee
are the mass eigenstates. Mixing in the left sector and 
the elimination of arbitrariness in the choice of the matrices
$V_d$ and $V_u$ can be considered in the same way as in the case of the SM.
The additional interaction Lagrangian of the right currents 
is expressed in terms of the mass eigenstates as follows:
\be
\sum_{i=1}^3 
\left( \bar U_i^R \hat W^+ D_i^R + \mbox{H.c.} \right) = 
\sum_{i=1}^3 
\bar u_i \gamma^\mu \hat W^+_R {1+\gamma^5 \over 2} d''_i + \mbox{H.c.},
\ee
where
\be
d_i''\;=\;Y_{ij} d_j,\ \ \ \ \ Y_{ij}\; =\; (Y^\dg_u)_{ik}(Y_d)_{kj}.
\ee
In the case of two flavors,
\bdm
Y=\left(
\begin{array}{cc}
\cos\theta_R & e^{i\delta_R}\sin\theta_R\\
 -\, e^{i\delta_R}\sin\theta_R & \cos\theta_R
\ea\right).
\edm
In the case of $N$ flavors the number of independent \CP-violating phases
is equal to ${N(N-1)\over 2}$.

\section{Low-Energy Effective Lagrangian}

\bc
\begin{figure*}[hbt] \hbox{
\hspace*{-5pt}
       \epsfxsize=350pt \epsfbox{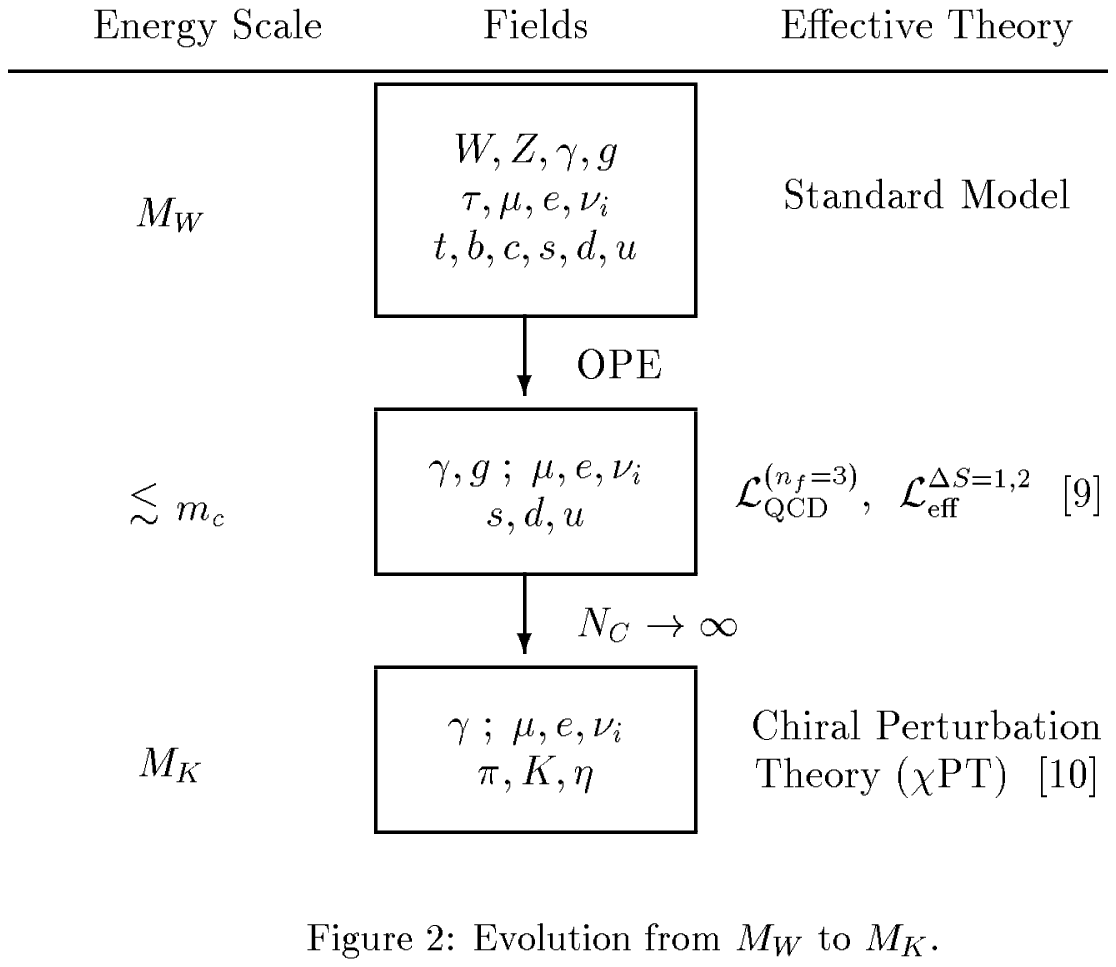} \hspace*{10pt}
       }
%
\end{figure*}
\ec
\setcounter{figure}{2}
\vspace*{-8mm}
It is well to recollect that any effective Lagrangian
derives from an expansion of the exact amplitudes at small external momenta.
Keeping only a few terms of such expansion (denote them by $T$), we find the 
Lagrangian $\cL_{eff}$ such that $T=\langle out |\cL_{eff}|in\rangle$.

Now we turn to the consideration of the consequences of the \CP-violating phase
for the hadronic physics. 
This can be made in the two stages:
\begin{itemize}
\item we construct the low-energy effective Lagrangian in terms of the quark fields,
where the CKM phase gives rise to the imaginary parts of the effective coupling constants,
represented by the Wilson coefficients;
\item using some model assumptions, we derive the expression for the Lagrangian in terms
of the meson fields.
\end{itemize}

\noindent As a result of the first stage of this evolution, 
one obtains the effective $\Delta S=1$ Lagrangian \cite{Okun}
\be\label{eq:EffQuarkLag}
 {\cal L}_{\mathrm eff}^{\Delta S=1}= - \frac{\GF}{\sqrt{2}}
 V_{ud}^{\phantom{*}}\,V^*_{us}\,  \sum_{i=1}^{10} C_i(\mu) \; Q_i (\mu) \; ,
\ee
which is a sum of local four--fermion operators $Q_i$,
constructed with the light degrees of freedom,
\bea
Q_{1} & = & \left( \overline{s}_{\alpha} u_{\beta}  \right)_{\rm V-A}
            \left( \overline{u}_{\beta}  d_{\alpha} \right)_{\rm V-A} \, , \\ \nonumber
Q_{2} & = & \left( \overline{s} u \right)_{\rm V-A}
            \left( \overline{u} d \right)_{\rm V-A} \, , \nonumber
\eea
where $(\bar q_i q_j)_{V\pm A}\equiv \bar q_i \gamma^\mu (1\pm\gamma^5) q_j$,
$\alpha$ and $\beta$ are color indices. The operators $Q_1$ and $Q_2$
and the respective coefficients $C_1$ and $C_2$ are determined by the low-energy
expansion of the diagrams in Fig.~3.
\bc
\begin{figure*}[thb] \hbox{
\hspace*{-5pt}
       \epsfxsize=420pt \epsfbox{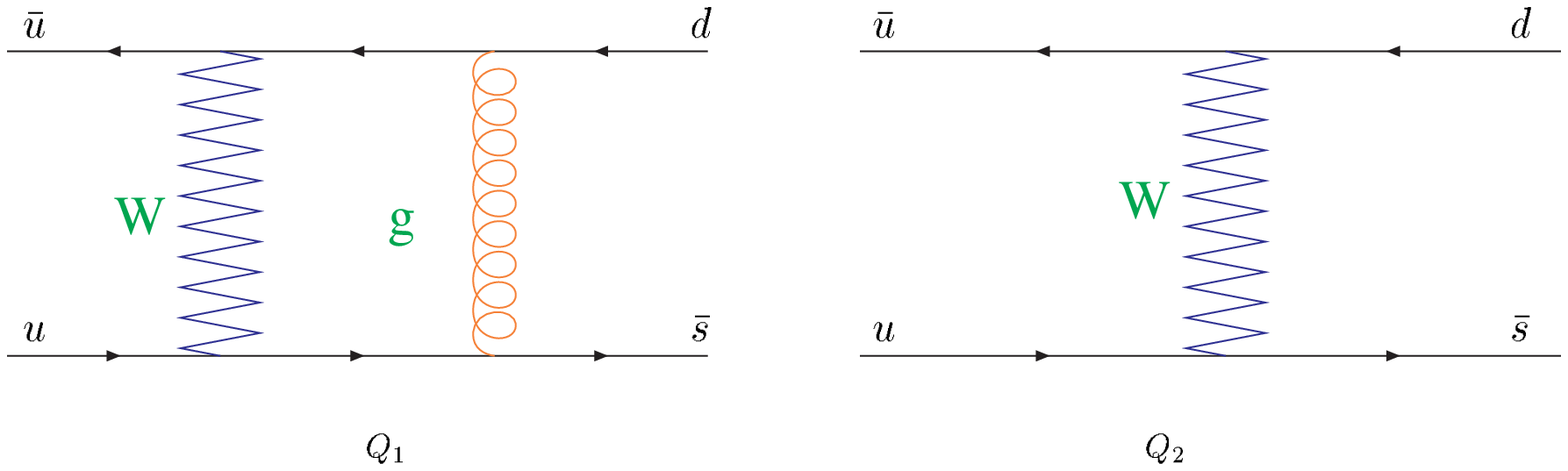} \hspace*{10pt}
       }
\caption{The diagrams for the short-distance processes giving rise to
the operators $Q_1$ and $Q_2$ in the effective Lagrangian (\ref{eq:EffQuarkLag}).} \label{fig:3}
\end{figure*}
\ec
\vspace*{-8mm}
The remaining operators 
\bea 
Q_{3,5} & = & \left( \overline{s} d \right)_{\rm V-A}
   \sum_{q} \left( \overline{q} q \right)_{\rm V\mp A} \, , \\ \nonumber
Q_{4,6} & = & \left( \overline{s}_{\alpha} d_{\beta}  \right)_{\rm V-A}
   \sum_{q} ( \overline{q}_{\beta}  q_{\alpha} )_{\rm V\mp A}\, , \\ \nonumber
Q_{7,9} & = & \frac{3}{2} \left( \overline{s} d \right)_{\rm V-A}
         \sum_{q} e_q \,\left( \overline{q} q \right)_{\rm V\pm A}\, , \\ \nonumber
Q_{8,10} & = & \frac{3}{2} \left( \overline{s}_{\alpha}
                                    d_{\beta} \right)_{\rm V-A}
  \sum_{q} e_q \,\left( \overline{q}_{\beta} q_{\alpha}\right)_{\rm V\pm A}\, ,\nonumber
\eea
come from the celebrated 'penguin' diagram:

\bc
\begin{figure*}[bbh] \hbox{
\hspace*{-5pt}
       \epsfxsize=400pt \epsfbox{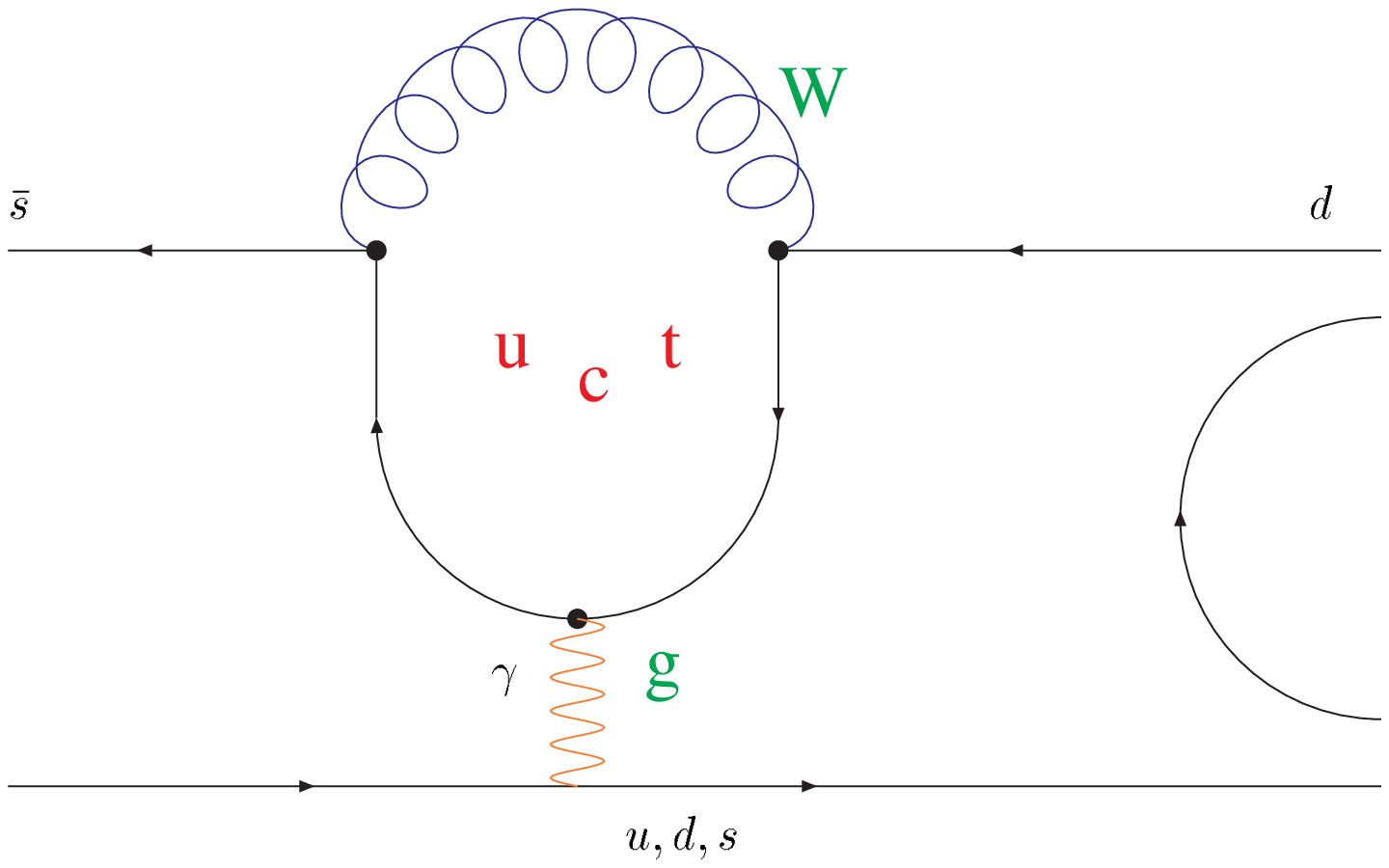} \hspace*{10pt}
       }
\caption{\bf The 'penguin' diagram.} \label{fig:4}
\end{figure*}
\ec
\pagebreak

\noindent The Wilson coefficients $C_i$ can be represented in the form
\be
C_i(\mu)=z_i(\mu)+\tau y_i(\mu),
\ee
where
\be
\tau = -\; {V_{td} V_{ts}^\ast \over V_{ud}V_{us}^\ast}.
\ee
The $\cC\!\cP$-violating amplitudes are proportional to $y_{i}$.

At the hadronic level, the effective Lagrangian is expressed in terms of the 
\begin{displaymath}
\mbox{meson fields:} \ \ \
 \ \Phi = \left(
\begin{array}{ccc}
\ds {\pi_0 \over \sqrt{2}} + {\eta_8 \over \sqrt{6}}+ {\eta_0 \over \sqrt{3}}
& \pi^+ & K^+ \\
\pi^- & \ds - {\pi_0 \over \sqrt{2}} + {\eta_8 \over \sqrt{6}} 
+ {\eta_0 \over \sqrt{3}} & K_0 \\
K^- & \bar K_0 &\ds - {2\eta_8 \over \sqrt{6}} + {\eta_0 \over \sqrt{3}}
\end{array}\right),
\end{displaymath}
The most general effective bosonic Lagrangian of the second order in derivatives, 
with the same $SU(3)_L\otimes SU(3)_R$ transformation properties and quantum numbers
as the short--distance Lagrangian, contains three terms \cite{Neufeld}:
\bea\label{eq:EffMesonLag}
\cL_2^{\Delta S=1} &=& -{G_F \over \sqrt{2}}\,  V_{ud}^{\phantom{*}} V_{us}^*
\; f^4\;\Bigg\{ g_8 \;\left[ \langle\lambda L_{\mu} L^{\mu}\rangle 
+ e^2 f^2 g_{ew}\;\langle\lambda U^\dagger \cQ U\rangle \right]
\Biggr. \nn\\ &&\qquad\qquad\qquad\quad\;\mbox{}  +
g_{27}\,\left( L_{\mu 23} L^\mu_{11} + {2\over 3} L_{\mu 21} L^\mu_{13}
\right)\Bigg\} \, ,
\eea
where the matrix 
\[
L_{\mu}=-i U^\dagger D_\mu U \ \ \ \ \left(U=\exp({i\sqrt{2}\Phi\over F_\pi})\right)
\]
represents the octet of
$V-A$ currents at lowest order in derivatives,
$\ds \cQ= {\rm diag}(\frac{2}{3},-\frac{1}{3},-\frac{1}{3})$ is the quark
charge matrix,
$\lambda\equiv (\lambda^6 - i \lambda^7)/2$ projects onto the
$\bar s\to \bar d$ transition [$\lambda_{ij} = \delta_{i3}\delta_{j2}$]
and $\langle {\mbox{A}}\rangle$ denotes the flavor trace of A.

The chiral couplings $g_8$ and $g_{27}$ measure the strength of the two
parts of the effective Lagrangian transforming as
$(8_L,1_R)$ and $(27_L,1_R)$, respectively, under chiral rotations.

In the presence of electroweak interactions, the explicit breaking
of chiral symmetry generated by the quark charge matrix $\cQ$ induces
the $\cO(p^0)$ operator $\langle\lambda U^\dagger \cQ U\rangle$, 
transforming as $(8_L,8_R)$ under the chiral group.
\be
\left| g_8 \right| \simeq 5.1\, , \qquad\qquad
\left| g_{27} \right| \simeq 0.29 \, .
\ee
The huge difference between these two couplings
shows the well--known enhancement of the octet $\vert\Delta I\vert = 1/2$ transitions.
In the $N_C\ra\infty$ limit, the real parts of these constants are expressed 
in terms of the Wilson coefficients as follows \cite{Bardeen}:
\bea\label{eq:c2}
g_8^\infty&=& -{2\over 5}\,C_1(\mu)+{3\over 5}\,C_2(\mu)+C_4(\mu)
- 16\, L_5\,\left( {\langle\bar q q \rangle^{(2)}(\mu) \over f^3}\right)^2
\,C_6(\mu)\, ,
\nonumber\\
g_{27}^\infty&=&{3\over 5}\,[C_1(\mu)+C_2(\mu)]\, ,
\\
(g_8\, e^2 g_{ew})^\infty&=& -3\,
\left( {\langle\bar q q \rangle^{(2)}(\mu) \over f^3}\right)^2
\, C_8(\mu)\, .
\nonumber\eea
The imaginary parts of them are responsible for \CP-violating effects
and will be considered further.
With the effective weak Lagrangian at hand, it is very helpful to consider 
the evolution of a meson system to the second order in the weak interaction.

\section{\CP\ Violation in System of Neutral Kaons}

It has become a tradition\footnote{In this Section, we follow
\cite{ChengLi, DAFNE, Nir}} to begin a description of the $K^0-\bar K^0$ system 
with writing down the most general expression for $\cC\cP\cT$-invariant Hamiltonian

\begin{displaymath}
i\; {d\over dt}
\left(\begin{array}{c} K^0 \\ 
		 \bar K^0 \\
	\end{array}\right)
= \left(
\begin{array}{cc}
H_{11} & H_{12} \\
H_{21} & H_{11} \\
\end{array}\right)
\left(\begin{array}{c}  K^0 \\ 
			\bar K^0 \\
	\end{array}\right),
\end{displaymath}
where $K^0$ and $\bar K^0$ are the eigenstates of the strong-interaction Hamiltonian
and the matrix $H_{ij}$ is non-Hermitean.

However, it is well to recollect a derivation of this formula and formulate
the assumptions made in its derivation.
\begin{itemize}
\item We consider weak interactions as a perturbation to the strong interactions.
\item We consider evolution of the eigenstates of the strong-interaction Hamiltonian 
to the second order in the weak interaction and then search for the
effective Hamiltonian that would give the same evolution in the leading order
of perturbation theory \cite{GellMann}.
\end{itemize}

The perturbation expansion of the $S$ matrix in the $K^0-\bar K^0$ system 
has the form
\be
S_{ab}=\langle b |T\,\exp\left(-i\int H^{int}_W(t)dt \right) |a\rangle = 
\delta_{ab} - 2\pi i T_{ab}.
\ee
where $a,b = K^0,\bar K^0$, $H^{int}_W(t)=e^{iHt}H_W e^{-iHt}$, and $H_W$
is the weak-interaction Hamiltonian. To the second order in $H_W$, we obtain
\bea
T_{ab}&=&\langle b |H_W|a\rangle -{i\over 2}\int dt
\langle b |T\,\left( H^{int}_W(t)  H^{int}_W(0) \right) |a\rangle  \nonumber \\
&=&\langle b |H_W|a\rangle +{1\over 2}\sum_\lambda
\left[ {\langle b |H_W | \lambda\rangle\langle\lambda| H_W |a\rangle \over E_b-E_\lambda+i\epsilon} \right.
+ \left. {\langle b |H_W | \lambda\rangle\langle\lambda| H_W |a\rangle \over E_a-E_\lambda+i\epsilon} \right]. 
\eea
Making use of the Sokhotsky relations one can represent the transition
amplitudes $T_{ab}+\langle b|H_{strong}|a\rangle$ as the matrix elements
of the effective Hamiltonian 
\be
H_{ab}=M_{ab}-i{\Gamma_{ab}\over2},
\ee
\vspace*{-2mm}where\vspace*{-2mm}
\bea\label{eq:GenFormeffCPVH}
M_{ab}&=& m_K \delta_{ab}+\langle b |H_W|a\rangle+{\cal P}\int d\lambda 
{\langle b |H_W | \lambda\rangle\langle\lambda| H_W |a\rangle \over m_K-E_\lambda}, \\ \nonumber
\Gamma_{ab}&=&2\pi\sum_\lambda \langle b |H_W | \lambda\rangle\langle\lambda| H_W |a\rangle \delta(m_K-E_\lambda).
\eea
Note that $M=M^\dagger$ and $\Gamma=\Gamma^\dagger$. Thus the Hamiltonian
\begin{displaymath}
H= \left(
\begin{array}{cc}
H_{11} & H_{12} \\
H_{21} & H_{11} \\
\end{array}\right) = \left(
\begin{array}{cc}
\ds (M'_{11}+iM''_{11})-i\,{(\Gamma'_{11}+i\,\Gamma''_{11})\over 2}; \ \  & 
\ds (M'_{12}-iM''_{12})-i\,{(\Gamma'_{12}-i\,\Gamma''_{12})\over 2} \\[4mm]
\ds (M'_{12}+iM''_{12})-i\,{(\Gamma'_{12}+i\,\Gamma''_{12})\over 2}; \ \ &
\ds (M'_{11}+iM''_{11})-i\,{(\Gamma'_{11}+i\,\Gamma''_{11})\over 2}  
\end{array}\right),
\end{displaymath}
is related to the Hamiltonian of the weak interactions.
Eigenstates of this Hamiltonian are identified with the physical states
$K^0_L$ and $K^0_S$, which are expressed through $K^0$ and $\bar K^0$
in terms of the parameter 
\be
\bar\epsilon={\sqrt{H_{12}}-\sqrt{H_{21}} \over \sqrt{H_{12}}+\sqrt{H_{21}}}:
\ee
\bdm
K^0_L={1\over\sqrt{2(1+|\bar\epsilon|^2)}}
\left(\begin{array}{c}  
	 1+\bar\epsilon\\ 
	 -1+\bar\epsilon\\
\ea\right),\hspace*{25mm}
K^0_S={1\over\sqrt{2(1+|\bar\epsilon|^2)}}
\left(\begin{array}{c}  
	 1+\bar\epsilon\\ 
	 1-\bar\epsilon\\
\ea\right);
\edm
the respective eigenvalues give the masses and widths
of the $K^0_L$ and $K^0_S$ mesons:
\bea\label{eq:Eigenvalues}
\lambda_S&=&H_{11}-\sqrt{H_{12}H_{21}}=M_S-i\,{\Gamma_S\over 2}, \\ \nonumber
\lambda_L&=&H_{11}+\sqrt{H_{12}H_{21}}=M_L-i\,{\Gamma_L\over 2}.  \nonumber
\eea

$\cC\!\cP$ transformation exchanges $|K^0\rangle$ and $|\bar K^0\rangle$ states:
\be
\cC\!\cP |K^0\rangle = e^{i\theta} |\bar K^0\rangle, \ \ \ \ 
\cC\!\cP |\bar K^0\rangle = e^{-\,i\theta} |K^0\rangle.
\ee
The phase factor can be chosen arbitrarily, because any quantum-mechanical state
is defined up to a phase. However \underline{\bf an interpretation} of the matrix
elements $H_{ab}$ and parameter $\bar\epsilon$ depends on a particular choice of 
the phase. 

In the case $\theta = 0$, the parameters $M''$ and $\Gamma''$ are $\cC\!\cP$-odd,
whereas $M'$ and $\Gamma'$ are $\cC\!\cP$-even. Assuming that $\cC\!\cP$-odd parameters are small,
\ $ \Gamma_{12}''<\!\!< \Gamma_{12}' \ \ \ \mbox{and} \ \ \ M_{12}''<\!\!< M_{12}'$, \
we obtain 
\be
\bar\epsilon = {H_{12}-H_{21}\over 4\sqrt{H_{12}H_{21}}+(\sqrt{H_{12}}-\sqrt{H_{21}})^2}
\approx {i\; M_{12}'' \over \lambda_S-\lambda_L}\; \mbox{sign}(M_{12}').
\ee
The formulas (\ref{eq:Eigenvalues}) agree with the experimental fact
\ \ $ M_{K_L} > M_{K_S} $ \ \ 
if the sign of the square root $\sqrt{H_{12}H_{21}}$ is chosen so that
\bea
\sqrt{H_{12}H_{21}} = M_{12}' - i\;{\Gamma'\over 2} \ & \mbox{for} &
M_{12}' > 0, \ \ \ \Delta\!M\equiv M_{K_S}-M_{K_L} = -2\,M_{12}';  \nonumber \\
\sqrt{H_{12}H_{21}} = - M_{12}' + i\;{\Gamma'\over 2} \ & \mbox{for} &
M_{12}' < 0, \ \ \ \Delta\!M\equiv M_{K_S}-M_{K_L} = 2\,M_{12}'. \nonumber
\eea
Experimental data indicate that $\ds \Delta M = -\; {\Delta \Gamma \over 2}$,
hence $\lambda_S - \lambda_L \approx \Delta M \; (1+i)$. Combining the above 
formulas, we arrive at
\be\label{eq:BasicEpsilon}
\bar\epsilon=-{e^{i\pi\over 4}\over \sqrt{2}}\; {M_{12}'' \over 2 M_{12}'}
\ee
for both $M_{12}'>0$ and $M_{12}'<0$. (The assumption that $M_{K_L}<M_{K_S}$
would give the phase factor $e^{-i\pi\over 4}$ instead of $e^{i\pi\over 4}$.)

\subsection{ Phase Convention}
It is important and helpful to keep track of the phase arbitrariness
stemming from the fact that 
\bdm
\mbox{both }\ \ 
\left( \begin{array}{c}
1 \\
0
\ea \right) \ \ 
\mbox{and} \ \ 
\left( \begin{array}{c}
e^{i\phi} \\
0
\ea \right) \ \ 
\mbox{describe the same physical state.}
\edm 
The transformation rules induced by the rotation of the phase of the $s$ quark
\be
s\ra e^{i\phi}s,\ \ \ \bar s\ra e^{-i\phi}\bar s,\
\ee
are as follows:
\[
K^0 \ra e^{-i\phi} K^0,\ \ \ \bar K^0 \ra e^{i\phi} \bar K^0,\ \ \ \ \ H_{12}\ra e^{-2i\phi} H_{12}, \ \ \ H_{21}\ra e^{2i\phi} H_{21},
\]
\[
 \ V_{us} \ra e^{i\phi} V_{us}  \ \ \ \ \ \ \ 
\bar\epsilon \ra {\bar\epsilon-i \,\mbox{tg}\phi \over 1 - i\,\bar\epsilon \mbox{tg}\phi}, \ \ \ A_I\ra e^{-i\phi} A_I,
\]
\bdm
\cC\!\cP=
\left(\begin{array}{cc}
0&1\\
1&0
\ea\right)
\ra
\left(\begin{array}{cc}
0&e^{-2i\phi} \\
e^{2i\phi}&0
\ea\right).
\edm
Let the phases of the $K^0$ and $\bar K^0$ be chosen so that the matrix of \CP\ transformation 
has the form 
\begin{displaymath}
\cC\!\cP_{ab}=\left(
\begin{array}{cc}
0 & 1 \\
1 & 0 \\
\end{array}\right)
\end{displaymath}
(we call it the "default" phase convention).
It should be compared with the widely used Wu--Yang phase convention,
in which the phase of $A_0$ is set equal to zero. It should be noticed that
in the Wu--Yang phase convention the operator of $\cC\!\cP$ transformation has the form
\begin{displaymath}
\cC\!\cP_{ab}=\left(
\begin{array}{cc}
0 & e^{-2i\alpha} \\
e^{2i\alpha} & 0 \\
\end{array}\right).
\end{displaymath}
It should also be emphasized that the value $\bar\epsilon$ as was 
defined above is phase-dependent and so {\bf does not measure \CP\ violation},
and the imaginary part of the effective weak Hamiltonian $H_W$ 
is not associated with $\cC\!\cP$ violation (and so it may be larger than real).

We adopt the "default" phase convention. In this case
\begin{itemize}
\item the quantities $M''$ and $\Gamma''$ are $\cC\!\cP$-odd; 
\item $M'$ and $\Gamma'$ are $\cC\!\cP$-even;
\item $V_{us}$ is real;
\end{itemize}

{\bf Upon fixing a phase convention,} the parameter $\bar\epsilon$ makes
physical sense and can be related to measurable quantities; the $\cC\!\cP$-violating parameters 
in the effective weak Hamiltonian are $M''$ and $\Gamma''$.

\subsection{Mixing of $K^0$ and $\bar K^0$ in the Standard Model}

As has been demonstrated, such mixing is accounted for by the $\Delta S=2$
effective weak Lagrangian. 

Let us consider the computation of the diagrams in Fig.~\ref{fig:5} (giving the 
transition amplitude $\bar s d \ra s\bar d$). 

Gaillard and Lee in the pioneer work \cite{GL1974} obtained 
\be
T(\bar s d \ra s\bar d) = -\;{\GF \over \sqrt{2}}\; {\alpha\over \pi \sin^2\theta_{\mbox{\tiny W}}}
(\bar d_L\gamma^\mu s_L)(\bar d_L\gamma_\mu s_L)\sum_{i,j=u,c,t} \xi_i\xi_j A(x_i,x_j),	\\[1mm]
\ee
\bc
\begin{figure*}[hhh] \hbox{
\hspace*{-5pt}
       \epsfxsize=450pt \epsfbox{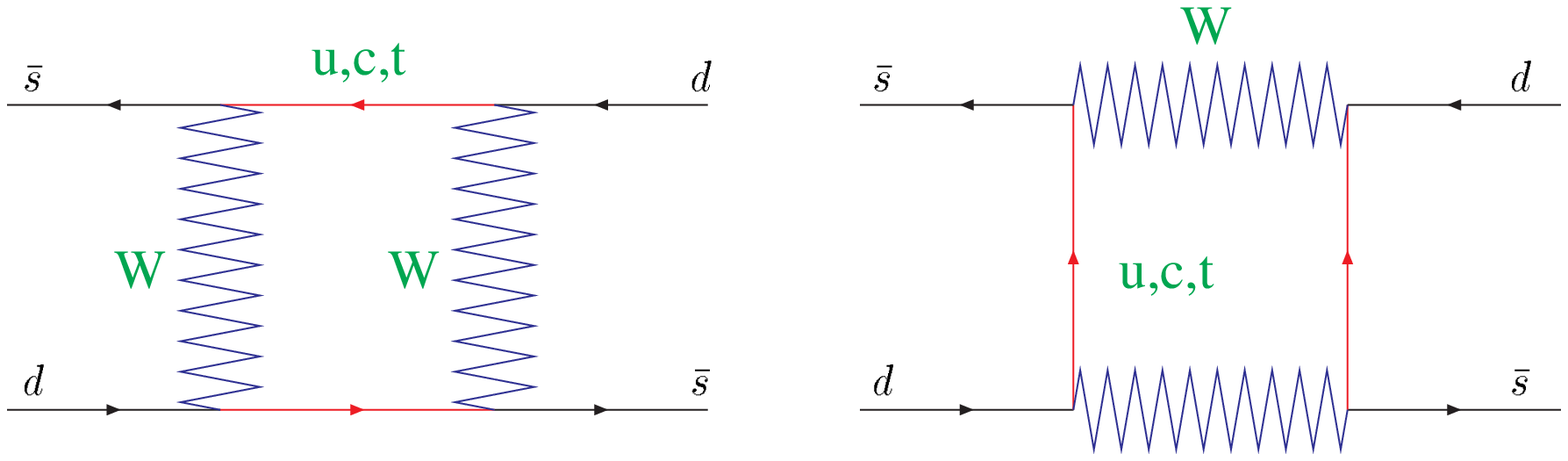} \hspace*{10pt}
       }
\caption{\bf $\Delta S=2$ transitions at the quark level.} \label{fig:5}
\end{figure*}
\ec
where 
\be
A(x_i,x_j)= {J(x_i)-J(x_j) \over x_i-x_j}, \ \ \ J(x) \equiv {1\over 1-x}\, + \, {x^2 \ln x \over (1-x)^2},
\ee
$$
\xi_i=V_{is}V_{id}^\ast, \ \ \ \mbox{ and } \ \ \ x_i={m_i^2 \over M_{\mbox{\tiny W}}^2}.
$$

Thus we obtain the effective $\Delta S=2$ Lagrangian\\[4mm]

\raisebox{15mm}{$\cL_{eff}^{\Delta s =2} \ \ = $}\hspace*{2mm}\begin{picture}(200,100)(0,0)
\SetColor{Magenta}
\ArrowLine(0,0)(100,50)
\ArrowLine(100,50)(0,100)
\ArrowLine(200,100)(100,50)
\ArrowLine(100,50)(200,0)
		\Text(15,16)[lb]{\large$d$}
		\Text(15,96)[lb]{\large$\bar s$}
		\Text(177,15)[lb]{\large$\bar d$}
		\Text(175,96)[lb]{\large$s$}
\SetColor{Brown}
\Vertex(100,50){8}
\end{picture}\hspace*{12mm}\raisebox{15mm}{$=-\; \ds
{\GF \over \sqrt{2}} {\alpha \over 16\pi \sin^2\theta_{\mbox{\tiny W}}} \; Q_0\; \lambda$,}\\[3mm]
where\footnote{To simplify these expression it is well to use the unitarity condition 
$\xi_u+\xi_c+\xi_t=0$.}
\bea
Q_0 &=& \bar d \gamma^\mu (1-\gamma^5) s\ \times \ \bar d \gamma_\mu (1-\gamma^5) s, \\[2mm] \nonumber
\lambda &=& \sum_{i,j} \xi_i \xi_j A(x_i,x_j). \\ \nonumber
\eea
With the use of this Lagrangian the mass difference is readily obtained:
\be
\Delta M = M_S-M_L\approx -2M_{12} = {1\over 2M_K}.
\langle \bar K^0 | \cL^{\Delta S=2}|K^0\rangle
\ee
Now one should evaluate the matrix element 
\be
\cM_{KK}=\langle \bar K^0| \bar d \gamma^\mu (1-\gamma^5) s\ 
\times \ \bar d \gamma_\mu (1-\gamma^5) s |K^0\rangle.
\ee
In early works, the matrix element was evaluated using the so called 
"Vacuum Insertion Approximation". The result is as follows:
\be 
\cM_{KK}={8\over 3} \langle \bar K^0| \bar d \gamma^\mu \gamma^5 s\ 
|K^0\rangle\ \langle \bar K^0|\bar d \gamma_\mu \gamma^5 s |K^0\rangle
= {8\over 3}M_K^2 F_K^2.
\ee
The first computation of this matrix element was performed in the bag model
\cite{Shrock}; in was found that it is smaller from the naive expectation
of $\cM_{KK}$ by a factor of 2. For this reason, the factor $B_K$ in the expression
\be
\langle \bar K^0|\ [ \bar d \gamma^\mu (1-\gamma^5) s]\ 
 \ [\bar s \gamma_\mu (1-\gamma^5) d]\ |K^0\rangle= {8\over 3}M_K^2 F_K^2\ B_K
\ee
is named "the bag constant". The computation of the bag factors presents 
the major challenge in the calculations of $\cC\!\cP$-violating quantities
in nonleptonic reactions.

The short-distance contribution to $\Delta M$ comprises $70\div 80\% $
of the total SM contribution:
\be\label{eq:DeltaMtheor}
\Delta M = \,-2\; M''_{12} = \,-\, {2\over 3} {\GF \over \sqrt{2}}\; \eta B_{K}\; {\alpha\over 4\pi}\,
\left(m_c \over 37 \mbox{GeV}\right)^2 
\re \lambda ,
\ee
where (in the approximation $x_u=0, x_c <\!\!< 1, x_t\sim 1$)\footnote{In the real
world, $x_u=2.5\times 10^{-11}, \ x_c=2\times 10^{-4}, \  x_t=4.7$.}
\be
\re \lambda = \re \left(\xi_c^2+\xi_t^2\; {x_t\over x_c}\; \right.
{(1-x_t^2+ 2 x_t \ln x_t)\over (1-x_t)^3 } + 2\xi_c\xi_t
\left. \left({x_t\over 1-x_t } + {\ln x_t\over (1-x_t)^2 } - \ln x_c\right)\right)
\ee
and the factor $\eta$ accounts for the corrections due to strong interactions, evaluated
in perturbative QCD.
The fact that $\re\lambda \approx 1$ indicates that the $c$ quark gives the dominant contribution
to the mass difference. The remaining $20\div 30\% $ are attributed to the long-distance
contribution (that is the contribution of the $\pi\pi$, $\pi\pi\pi$ {\it etc} 
intermediate states in formula (\ref{eq:GenFormeffCPVH})), 
which is extremely difficult to compute exactly.
\bc
\begin{figure*}[hhh] \hbox{
\hspace*{-5pt}
       \epsfxsize=400pt \epsfbox{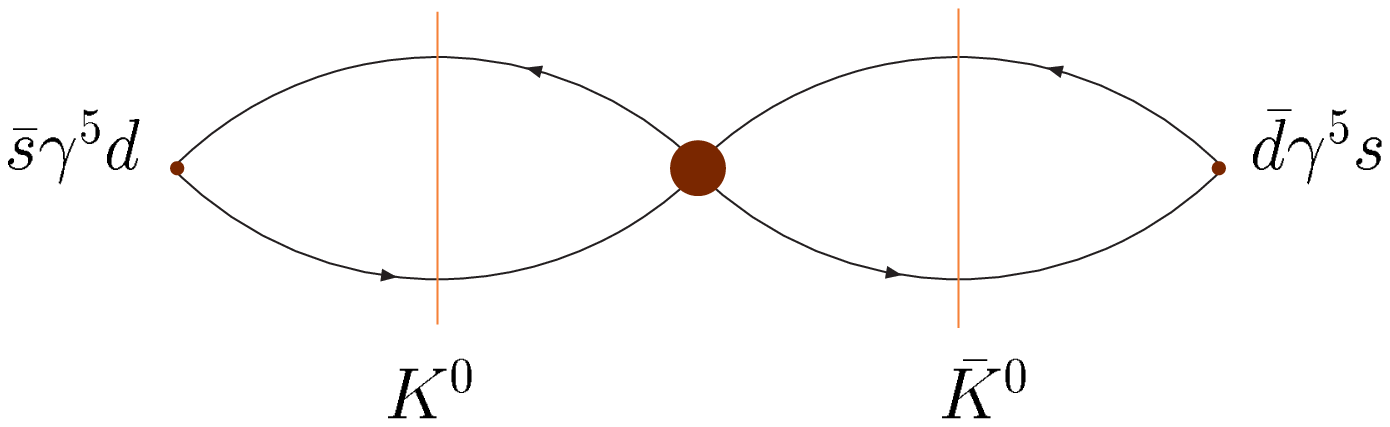} \hspace*{10pt}
       }
\caption{\bf The $K^0-\bar K^0$ transition in terms of the effective Lagrangian.} \label{fig:6}
\end{figure*}
\ec
The estimates of the bag constant obtained in the lattice QCD and in
some models are 
\bea
B_K&=&0.85\pm 0.15 \ \ \ (\mbox{Lattice\ \ \cite{Lellouch}}),  \\
B_K&=&0.41\pm 0.09 \ \ \ (\mbox{Chiral limit}, 1/N_C \ \ \cite{Cuichini}).
\eea

\subsection{Basic Formula for $\epsilon$}
In the above subsection we have considered in detail the determination of the real part 
of the amplitude of the $\Delta S=2$ transition in terms of
the short-distance contribution (the second term in the expression (\ref{eq:GenFormeffCPVH})
for $M_{ab}$) and the bag constant $B_K$. 
The imaginary part of this amplitude,
which appears in the expression (\ref{eq:BasicEpsilon}) for $\epsilon$,
can be calculated by the same token.
In the case of imaginary part, one can safely neglect the long-distance contribution due
to low-lying intermediate states associated with the third 
term in the expression (\ref{eq:GenFormeffCPVH}) for $M_{ab}$).
The result is
\be\label{eq:EpsilonTheor}
\epsilon_{\mbox{teor}} = C_\epsilon B_K \exp\left({i\pi\over 4}\right)\; \im \xi_t
\left[\re \xi_c\left(\eta_1 S_0(x_c)-\eta_3 S_0(x_c, x_t)\right) - \re \xi_t \eta_2 S_0(x_t)\right],
\ee
where 
\be
C_\epsilon = {\GF^2 F_K^2 \, M_K \, M_W^2 \over 6\sqrt{2} \pi^2 \, \Delta\!M_K} = 3.837\times 10^4
\ee
(here we have used the experimental value of $M'_{12}$, determined from $\Delta M_K$,
instead of the theoretical value (\ref{eq:DeltaMtheor})),
\bea
S_0(x_c) &\approx& x_c, \\ \nonumber
S_0(x_t) &\approx& 2.46\cdot \left( m_t\over 170 \mbox{GeV}\right)^{1.52}, \\ \nonumber
S_0(x_c,x_t) &\approx& x_c \left[\ln\left(x_t\over x_c\right) \,-\, {3x_t\over 4(1-x_t)}\,-\, {3 x_t^2 \ln x_t \over 4(1-x_t)^2}\right], \\ \nonumber
\eea
\[
m_c\simeq 1.3 \mbox{GeV}, \ \ \  m_t=174.3\pm 5.1 \mbox{GeV};
\]
the short-distance corrections due to the strong interactions are absorbed
in the coefficients \cite{Herrlich}
\[
\eta_1=1.38\pm0.20, \ \ \ \  \eta_2=0.57\pm 0.01, \ \ \ \ \ \eta_3=0.45\pm 0.04.
\]
Formula (\ref{eq:EpsilonTheor}) allows to set a limitation on the 
\CP-violating parameter $\eta$ of the SM from the experimental 
limitations on $\epsilon$ (see Fig.(\ref{fig:7})).
\bc
\begin{figure*}[hhh] \hbox{
\hspace*{-5pt}
       \epsfxsize=340pt \epsfbox{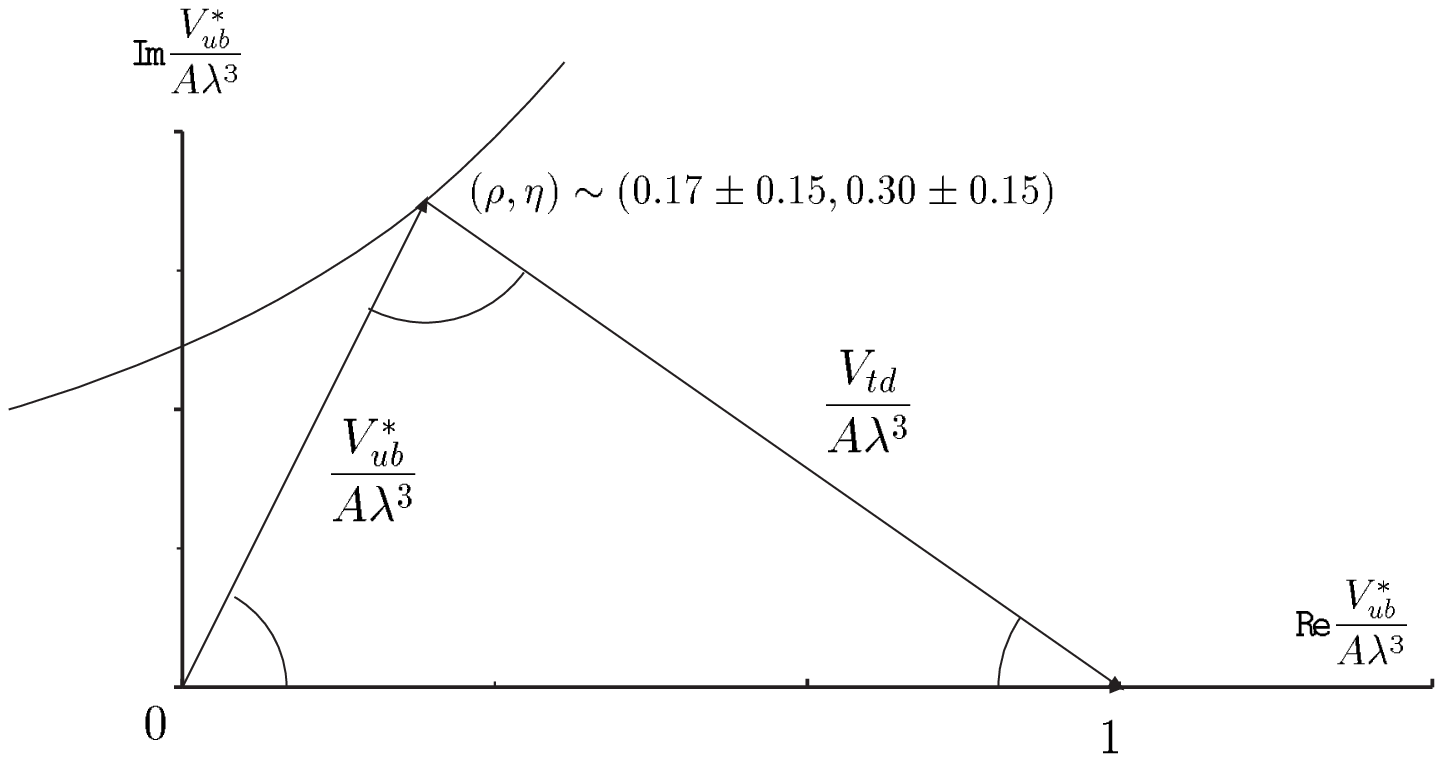} \hspace*{10pt}
       }
\caption{Formula (\ref{eq:EpsilonTheor}) confines $\epsilon$ to lie in some vicinity of
the indicated hyperbola (see \cite{PDG}). The combined set of 
limitations makes the vertex of the unitarity triangle to lie within the indicated limits.} \label{fig:7}
\end{figure*}
\ec

\subsection{ ($K\ra\pi\pi$) Amplitudes}
The $2\pi$ decay amplitudes of the neutral Kaons in the channel with isospin $I$
are defined by the matrix elements
\bea
\langle \pi\pi,\; I|H_W|K^0\rangle &=& 
\sqrt{3\over 2} (A_I+B_I)\, e^{i\delta_I}, \\ \nonumber
\langle \pi\pi,\; I|H_W|\bar K^0\rangle &=& 
\sqrt{3\over 2} (A_I^\ast-B_I^\ast)\, e^{i\delta_I}, \nonumber
\eea
where $\delta_I$ is the $S$-wave phase shift of the $\pi\pi$ scattering;
\be
|\pi\pi,\; I\; \mbox{out} \rangle = e^{2i\delta_I} |\pi\pi,\; I\; \mbox{in} \rangle
\ee
and
\be 
A_I=A_I'+iA_I'', \ \ \ \ B_I=B_I'+iB_I''.
\ee
The properties of these amplitudes under the \CP-transformation and 
time reversal $\cT$ are seen from Table 2.\\[1mm]

{\bf Table 2.} Transformation properties of the isospin amplitudes.\\[2mm]
\hspace*{23mm}\begin{tabular}{|c|c|c|c|c|}
\hline
&&&& \\
Transformation& $\ \ A'\ \ $ & \ \ $A''$ \ \  & $\ \ B'\ \ $ & $\ \ B''\ \ $ \\
&&& &  \\ \hline
&&&& \\
\CP & + & --- & ---  & + \\
&&& &  \\ \hline
&&&& \\
$\cT$ & + & --- & + & --- \\
&&& &  \\ \hline
\end{tabular}
\vspace*{2mm}

\noindent It is seen that the amplitudes $A_I$ are $\cC\cP\cT$-even, whereas the $B_I$ amplitudes
are $\cC\cP\cT$-odd. 

Using the relations 
\bea
 {1\over \sqrt{2}}|\pi^+\pi^- + \pi^-\pi^+\rangle
&=& {1\over \sqrt{3}}(\sqrt{2}|I=0\rangle + |I=2\rangle), \nonumber \\
 |\pi^0\pi^0 \rangle &=& {1\over \sqrt{3}}(|I=0\rangle -\sqrt{2} |I=2\rangle), \nonumber
\eea
we obtain the expressions for the amplitudes $\cA(K^0\ra\pi^+\pi^-)$ 
and $\cA(K^0\ra\pi^0\pi^0)$:
\bea
\cA(K^0\ra\pi^+\pi^-) &=& A_0 e^{i\delta_0} + {1\over \sqrt{2}} A_2 e^{i\delta_2}, \\ \nonumber
\cA(K^0\ra\pi^0\pi^0) &=& {1\over \sqrt{2}} A_0 e^{i\delta_0} - A_2 e^{i\delta_2}. \nonumber
\eea
$A_0$ describes the transitions with $\Delta I = 1/2$, whereas 
$A_2$ describes the transitions with $\Delta I = 3/2$. 
Assuming that the Hamiltonian for the transitions $K\ra 2\pi$
contains only terms with quantum numbers $I=1/2, I_3=1/2$ and $I=3/2, I_3=1/2$
(that is, it does not contain the terms with $I=5/2, I_3=1/2$ {\it etc.}),
we can use the Clebsch--Gordan coefficients
\bdm
\begin{array}{rcl}
&&\\[3mm]
\langle \pi^+\pi^0 |&H_W&|K^+\rangle \\[3mm]
&&\\[3mm]
\langle \pi\pi, I=2 |&H_W&|K^+\rangle 
\ea
\begin{array}{cccccc}
&\langle J, M | &j_1, m_1;&| j_2, m_2 \rangle && \\[3mm]
\sim&\langle 2,1 |&  {3\over 2}, {1\over 2}; &  {1\over 2}, {1\over 2}\rangle&=&\sqrt{3}/2 \\[3mm]
&\langle \pi\pi | & H_W;&| K \rangle && \\[3mm]
\sim&\langle 2,0 |&  {3\over 2}, {1\over 2}; &  {1\over 2}, -\,{1\over 2}\rangle&=&1/\sqrt{2} 
\ea
\edm
to obtain the $\cA(K^+\ra\pi^+\pi^0)$ amplitude
\be
\cA(K^+\ra\pi^+\pi^0) = {3\over 2} A_2 e^{i\delta_2}.
\ee
A comparison of the lifetimes of $K^0_S$ and $K^+$ leads one to a conclusion that the ratio
\be\label{eq:omega}
|\omega| \;=\;{|A_2| \over |A_0|} \;=\; {1\over 22}
\ee
is very small, which is referred to as the long-standing "$\Delta I = 1/2$ problem".
The experimental value $\delta_2-\delta_0 \simeq (45\pm 6)^o$.

The tree--level $K\to\pi\pi$ amplitudes generated
by the $\cO(p^2)$ $\chi$PT Lagrangian are:
\bea
\label{eq:AMP_2}
A_0 &\! =&\! -{G_F\over \sqrt{2}} \, V_{ud}^{\phantom{*}} V_{us}^*\;\sqrt{2}\,
f\;\left\{\left ( g_8+{1\over 9}\, g_{27}\right )(M_K^2-M_\pi^2)
-{2\over 3}\, f^2\, e^2 \; g_8\; g_{ew}\right\}\; ,
\nonumber\\
A_2&\! =&\!  -{G_F\over \sqrt{2}} \, V_{ud}^{\phantom{*}} V_{us}^*\; {2\over 9}
\,f\;\biggl\{ 5\, g_{27}\, (M_K^2-M_\pi^2) -3\, f^2\, e^2\; g_8\; g_{ew}
\biggr\}
\; .
\eea

Let us introduce the notation 
\be
\omega\; = \; {A_2 \over A_0} \;=\; |\omega| e^{i\chi},\ \ 
A_0=|A_0| e^{i\alpha}, \ \ 
\delta = \delta_2 - \delta_0, \ \ 
\tilde \omega = \sqrt{2}|\omega| e^{i\delta}.
\ee
We can now express the amplitudes $\cA(K^0_L\ra\pi^+\pi^-)$, $\cA(K^0_L\ra\pi^0\pi^0)$
$\cA(K^0_S \ra\pi^+\pi^-)$, and $\cA(K^0_S \ra\pi^0\pi^0)$ in terms of $A_1, A_2$ and the 
parameter $\bar\epsilon$. For example,
\bea
\cA(K^0_L\ra\pi^0\pi^0) &=& {1\over \sqrt{2(1+|\bar\epsilon|^2)}}
\{ [ \cA(K^0\ra\pi^0\pi^0) - \cA(\bar K^0\ra\pi^0\pi^0)] \\ \nonumber
&&\hspace{23mm} + \bar \epsilon  [ \cA(K^0\ra\pi^0\pi^0) + \cA(\bar K^0\ra\pi^0\pi^0)] \}\\ \nonumber
&=&{|A_0|e^{i\delta_0} \over \sqrt{2(1+|\bar\epsilon|^2)}}
\left( i\,\sin\alpha -i\tilde \omega \sin(\alpha+\chi) + \bar\epsilon [\cos\alpha - \tilde \omega \cos(\alpha+\chi)] \right)
\eea
Let us consider the experimentally measurable values
\be
\eta_{00}=\etazz, \ \ 
\eta_{+-}=\etapm.
\ee
Proceeding as indicated above, we obtain
\be
\eta_{00}={i\,[\sin\alpha - \tilde \omega \sin(\alpha+\chi)] + \bar\epsilon [\cos\alpha - \tilde \omega \cos(\alpha+\chi)]
\over [\cos\alpha - \tilde \omega \cos(\alpha+\chi)] + i\bar\epsilon [\sin\alpha - \tilde \omega \sin(\alpha+\chi)] }
\ee
and a similar expression for $\eta_{+-}$.

Assuming that $\cC\!\cP$-violating parameters are small, 
that is $\bar\epsilon, \alpha, \chi <\!\!< 1$,
we arrive at
\be
\eta_{00}=\bar\epsilon +i \alpha- i\,{ \chi\tilde\omega \over 1-\tilde\omega},
\ee
where 
\[
\chi \approx {A''_2 \over A'_2}\; -\; {A''_0 \over A'_0}.
\]
In a similar way, one obtains
\be
\eta_{+-}\equiv \etapm =\bar\epsilon +i \alpha- i\,{ \chi\tilde\omega \over 1-\tilde\omega}.
\ee

Let us introduce the parameters, which are conventionally used for a description of the
effects of $\cC\!\cP$ violation:\\[1mm]
\bea
\epsilon &=& {{\langle \pi\pi, I=0 | K^0_L \rangle \over \langle \pi\pi, I=0 | K^0_S \rangle}}
= \bar\epsilon + i\alpha, \\[2mm] \nonumber 
\epsilon'&=& {i \over \sqrt{2}} e^{i\delta} Im{A_2\over A_0} = 
{i\;\tilde\omega \over 2}\left({A''_2 \over A'_2}\; -\; {A''_0 \over A'_0}\right).\\ \nonumber 
\eea
Note that, in contrast to $\bar\epsilon$, the parameter $\epsilon$ is independent of
the phase convention. These values coincide {\bf only in the Wu--Yang phase convention.}
In terms of the introduced parameters, we have
\bea\ds
\eta_{00}&=&\etazz=\epsilon - {2\epsilon' \over 1-|\omega|e^{i\delta}\sqrt{2}}\approx \epsilon -2\epsilon' \\ \nonumber
\eta_{+-}&=&\etapm=\epsilon + { \epsilon' \over  1+{|\omega|e^{i\delta}\over\sqrt{2}}}\approx \epsilon + \epsilon', \\ \nonumber
\eea
where
\[
|\omega|=Re{A_2\over A_0}\simeq 0.045,\ \ \  \delta\simeq 45^o.
\]
From the above it follows that
\be
Re{\epsilon'\over \epsilon}\simeq {1-|\omega|\over 6}\left(1-\left|{\eta_{00}\over \eta_{\pm}}\right|^2\right).
\ee
Note that $\epsilon'/ \epsilon$ is approximately real.
Using the short--distance Lagrangian, the \CP--violating ratio
$\epsilon'/\epsilon$ can be written as follows \cite{Buras}:
\be\label{eq:EpsEpsPrimPich}
{\epsilon^\prime\over\epsilon} \, = \,
\mbox{Im}\left(V_{ts}^* V_{td}^{\phantom{*}}\right)\, e^{i\Phi}\; 
{G_F\over 2 \vert\varepsilon\vert}\; {\omega\over |\mbox{Re}(A_0)|}\;
\left [P^{(0)}\, (1-\Omega_{IB}) - {1\over \omega} \,P^{(2)}\right ]\, ,
\ee
where the quantities
\be
P^{(I)}= \sum_i\, y_i(\mu)\;
\langle (\pi\pi )_I\vert Q_i\vert K\rangle 
\ee
contain the contributions from hadronic matrix elements 
with isospin $I$ \ and
\be
\Omega_{IB} = {1\over \omega}\,
{\mbox{Im}(A_2)_{\mbox{\small{IB}}}  \over \mbox{Im}(A_0)}
\ee
parameterizes isospin breaking corrections.
The factor $1/\omega$ enhances the relative weight of the
$I=2$ contributions.

The hadronic matrix elements 
$\langle (\pi\pi )_I\vert Q_i\vert K\rangle $
are usually parameterized
in terms of the bag parameters $B_i$, which measure them
in units of their vacuum insertion approximation values.
In the SM, $P^{(0)}$ and $P^{(2)}$ turn out to
be dominated by the contributions from 
the QCD penguin operator $Q_6$ and
the electroweak penguin operator $Q_8$, respectively \cite{trieste}.
Thus, to a very good approximation,
$\epsilon'/\epsilon$ can be written (up to global factors)
as \cite{munich, Buras93}
\be
{\epsilon'\over\epsilon} \,\sim\,
\left [ B_6^{(1/2)}(1-\Omega_{IB}) - 0.4 \, B_8^{(3/2)}
 \right ]\, .
\ee

The isospin--breaking correction coming from $\pi^0$-$\eta$  mixing
was originally estimated to be $\Omega_{IB}^{\pi^0\eta}=0.25$ 
\cite{Donoghue86}. Together with the usual ansatz $B_i\sim 1$, 
this produces a large numerical cancellation
in (\ref{eq:EpsEpsPrimPich}) \cite{Pich} leading to low values of $\epsilon'/\epsilon$
around $7\cdot 10^{-4}$.
A recent improved calculation of $\pi^0$-$\eta$  mixing
at $\cO(p^4)$ in $\chi$PT has found the result \cite{Ecker00}
\be
\Omega_{IB}^{\pi^0\eta}\, =\, 0.16\pm 0.03 \, .
\ee
This smaller number, slightly increases the naive estimate of $\epsilon'/\epsilon$.\\[2mm]

{\bf Table 3.} Theoretical and experimental values of $\epsilon'/\epsilon$.\\[2mm]

\hspace*{-4mm}\begin{tabular}{|c|c|c|c|}
\hline
&&& \\
Year & Theory (models: & Theory & Experiment  \\
&  $1/N_c$, unitarization) & (lattice) &   \\ \hline

&&& \\
$\sim$ 1988 & $0.01\div 0.03$ \ \cite{ChengLi} & & $(3.2\pm 1.0) \times 10^{-3}$ \cite{PDG88} \\
 &  & &   \\
&&&   \\ \hline

&&& \\
1995 & $(6.7\pm 2.6) \times 10^{-4}$\cite{Buras93}  & $(3.1\pm 2.5) \times 10^{-4}$\cite{DAFNECiuchini} & $(1.5\pm 0.8) \times 10^{-3}$ \cite{PDG96} \\
 &  & &   \\
&&&   \\ \hline

&&& \\
1999 & $(-1\div 35)\times 10^{-4}$ \ \ \cite{Buras} &$(0.44\div 2.1) \times 10^{-3}$ \ \ 
\cite{Aoki}$^{\ast}$ & $(2.1\pm 1.5) \times 10^{-3}$ \cite{PDG}  \\
--2000 &  & &   \\
&&&   \\ \hline

&&& \\
2001 & $(1.7\pm 0.9) \times 10^{-3}$ \ \cite{Pich} &  & $(1.53\pm 0.24) \times 10^{-3}$ \ 
\cite{NA48}$^{\ast\ast}$ \\
 &  & &  \\
&& &  \\ \hline
\end{tabular}

\vspace*{3mm}
$^{\ast}$ {\small This value depends crucially on the mass of the $s$ quark:
$0.44\times 10^{-3}$ for $m_s(m_c)=150$~MeV and $2.1\times 10^{-3}$ for $m_s(m_c)=80$~MeV}

$^{\ast\ast}$ {\small The most recent data: $\re {\epsilon'\over \epsilon} = (1.73\pm 0.18) 
\times 10^{-3}$ \\
(http://na48.web.cern.ch/NA48/Welcome/images/talks/win02/win02.pdf)}
 
\newpage
\section{Time-Reversal Invariance}

Throughout this Section it is assumed that all the processes under consideration
are adequately described by some local quantum field theory, that is, 
$\cC\cP\cT$ is an exact symmetry of the theory. In this case \CP-violation
is equivalent to $\cT$-violation and so we turn to the consideration of the
time-reversal invariance.
\vspace*{3mm}
\bc
\begin{figure*}[hhh] \hbox{
\hspace*{-5pt}
       \epsfxsize=420pt \epsfbox{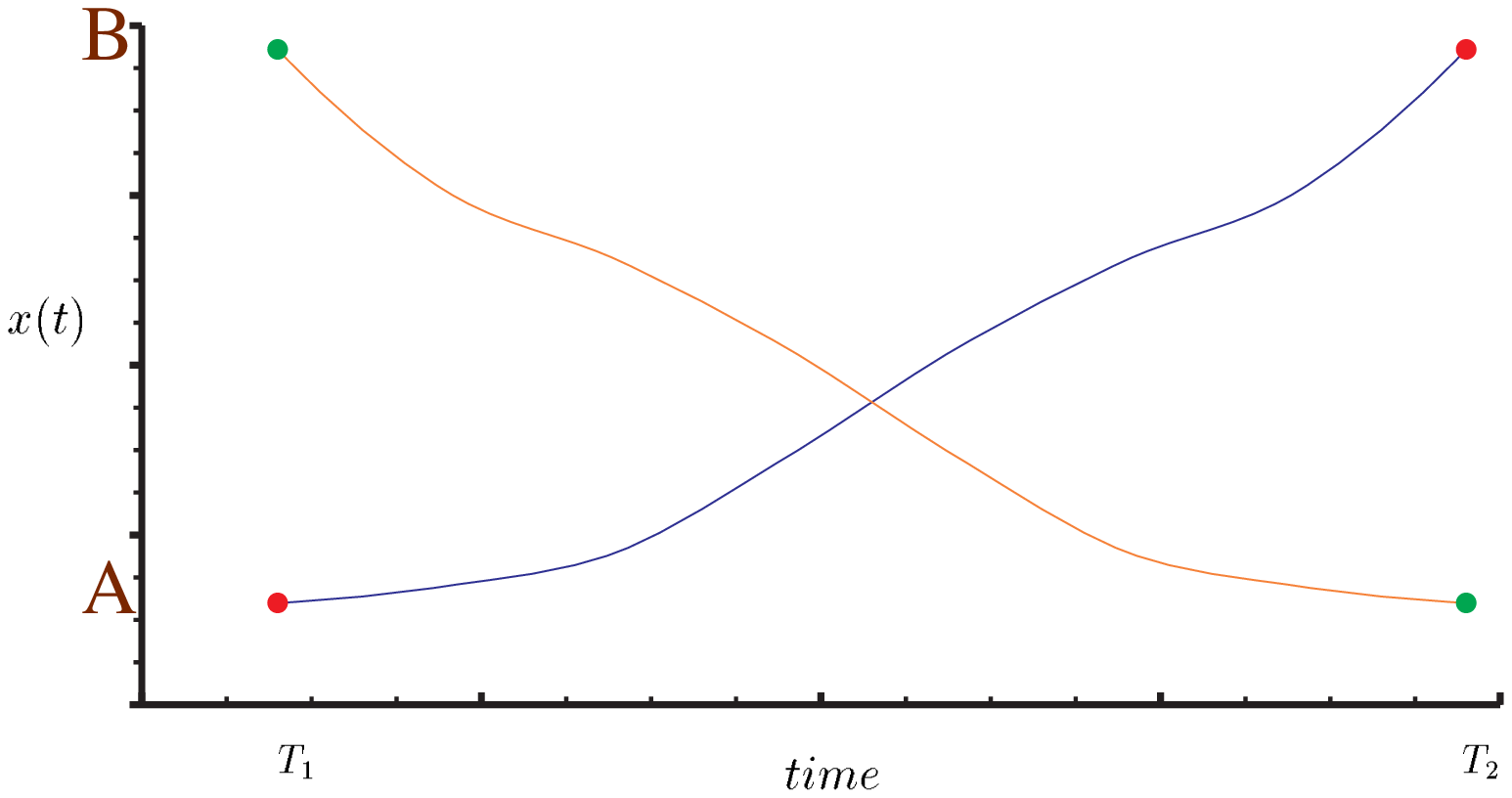} \hspace*{10pt} 
       }
\caption{\bf The reversal of time in classical mechanics.} \label{fig:8}
\end{figure*}
\ec

\underline{Time-reversal invariance in the classical mechanics} (Fig.~\ref{fig:8}): \\[2mm]
motion from $A$ to $B$ $\ \ \ \{x(t): \ x(T_1)=A, \ x(T_2)=B\}$ \\[1mm]
\centerline{and }
motion from $B$ to $A$ $\ \ \ \{x(t): \ x(T_1)=B, \ x(T_2)=A\}$ \\[3mm]
\centerline{\bf are described with the same Hamiltonian}\\
This is the case provided that $H(p,x)=H(-p,x)$.\\[4mm]

\underline{Time-reversal invariance in the quantum mechanics}: \\[2mm]
evolution from $|A\rangle$ to $\langle B |$ $\ \ \ \{|\psi(t)\rangle: \ |\psi(T_1)=|A\rangle, \ \langle \psi(T_2)|=\langle B|\}$ \\[1mm]
\centerline{and }
evolution from $|B\rangle$ to $\langle A| $ $\ \ \ \{|\psi(t)\rangle: \ |\psi(T_1)\rangle =|B\rangle, \ \langle (T_2)|=\langle A|\}$ \\[3mm]
\centerline{\bf are described with the same Hamiltonian}
This is the case provided that $H=H^\ast$.\\[4mm]
\newpage

\underline{{\bf Table 4.} $\cP$ and $\cT$ transformations for various quantities.}\\
Action of these operators on the quantum-mechanical states is determined by the condition that
the $\cT$- (or $\cP$-)transformed states are characterized by the $\cT$- (or $\cP$-)transformed
(eigen)values of the respective operators.\\[4mm]

\hspace*{-3mm}\begin{tabular}{|l|c|c|c|l|}
\hline
&&&& \\
Value & Notation & $\cP$-transformed  & $\cT$-transformed & Comment \\
&&value&value &  \\ \hline

&&&& \\
Coordinate & {$\bf \vec x$} &{$\bf -\ \vec x$}&{$\bf \vec x$}&  \\
&&&& \\ \hline

&&&& \\
Momentum &{$\bf \vec p$} &{$\bf -\ \vec p$}&{$\bf-\ \vec p$}&   $\displaystyle {\bf \vec p} =m\;{d {\bf \vec x} \over d t}$\\
&&&& \\ \hline

&&&&  \\
Angular momentum &{$\bf \vec l$} &{$\bf \vec l$}&{$\bf -\ \vec l$}&  ($\bf \vec r\times \vec p$) \\
&&&& \\ \hline

&&&& \\
Spin &${\bf \vec s}$ &{$\bf \vec s$}&{$\bf -\ \vec s$}& Like \ \ $\bf \vec l$ \\
&&&& \\ \hline

&&&& \\
Electric field &{$\bf \vec E$} &{$\bf -\ \vec E$}&{$\bf  \vec E$}&  $ \displaystyle {\bf \vec E} =-\;{d  A_0 \over d {\bf \vec r}}$ \\
&&&& \\ \hline

&&&& \\
Magnetic field &{$\bf \vec B$} &{$\bf \vec B$}&{$\bf -\ \vec B$}&  $ {\bf \vec B} \sim \vec r \times \vec j$ \\
&&&& \\ \hline

&&&& \\
Potential &{$A_0$} & $ A_0$ &$A_0$ &    \\
&&&& \\ \hline

&&&& \\
Vector potential &{$\vec A$} & $-\; \vec A$ &$-\vec A$&    \\
&&&& \\ \hline

&&&& \\
Helicity & $\lambda$ &  $-\; \lambda$ &  $\lambda$ &  $\lambda = {\bf \vec s \vec p}$ \\
&&&& \\ \hline

&&&& \\
Transverse polarization$^{\ast}$ & $\xi$ & $\xi$ & $-\; \xi$ & $\xi = {\bf (\vec s, [\vec p_1 \times \vec p_2])}$   \\
&&&& \\ \hline

&&&& \\
Triple correlation$^{\ast\ast}$ & $\eta$ & $-\; \eta$ & $-\; \eta$ & $\eta=\bf (\vec p_1, [\vec p_2 \times \vec p_3])$  \\
&&&& \\ \hline

\end{tabular}

\vspace*{3mm}

$^{\ast}$ {\small A characteristic of a three-particle state, if at least one particle
has a nonzero spin.}

$^{\ast\ast}$ {\small A characteristic of a multiparticle state (number of particles
must be greater than 3).}


\newpage
\bc
\begin{figure*} \hbox{
\hspace*{-5pt}
       \epsfxsize=420pt \epsfbox{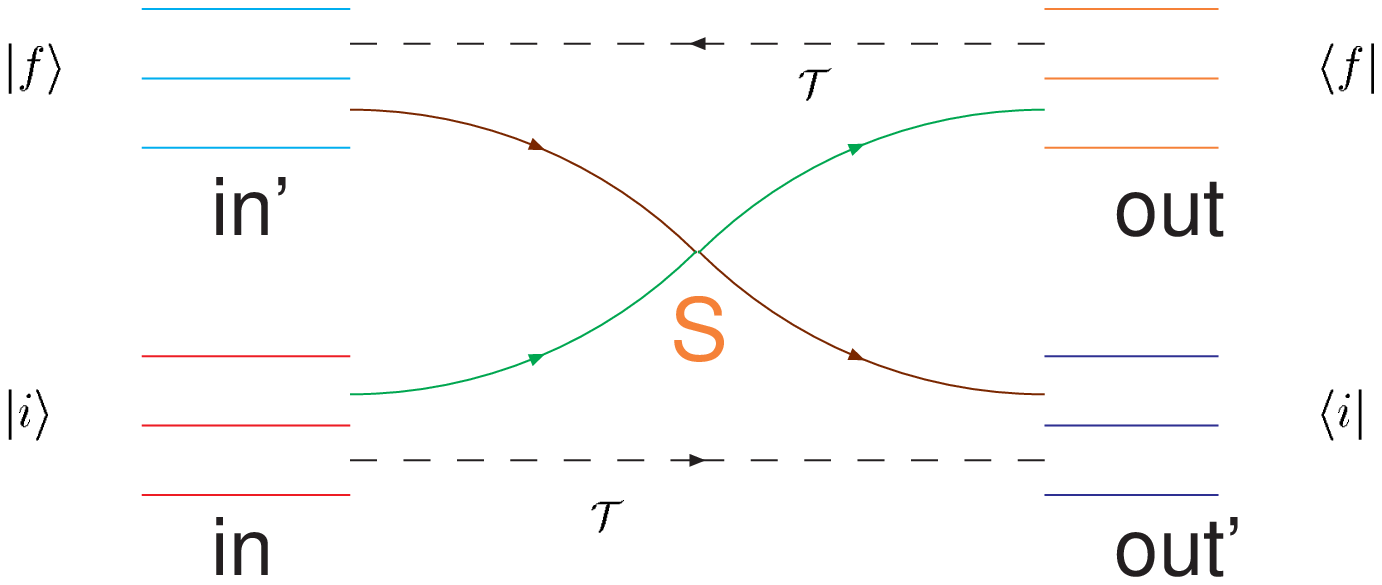} \hspace*{10pt}
       }
\caption{\bf The reversal of time in the $S$-matrix approach.} \label{fig:9}
\end{figure*}
\ec

\underline{Time-reversal invariance in the $S$-matrix approach}: \\[2mm]

The system described by the quantum field theory is invariant under
the time reversal if
\begin{itemize}
\item the space of $in$-states ('ket'-vectors) is isomorphic to the 
space of $out$ states ('bra'-vectors)\footnote{This isomorphism is nothing
but assumption; however, it allows to consider the vectors from $in$ and $out$ 
spaces as the vectors of the same Hilbert space $\cH$,
identified with $in$.}: $\forall | I \rangle \in in\ \ \exists\ \langle I | \in out$;
\item There exists (anti-unitary) operator of time reversal
\[\cT: in\ra out,\ \ \cT: out\ra in, \ \ \mbox{such that}\ \ \
\langle \cT I | \cT F \rangle = \langle F | I \rangle= \langle I | F \rangle^\ast,
\]
which changes the signs of spins and momenta of all particles;
this condition can be cast in the form 
$(\phi,\;\cT \psi) = (\cT\phi,\; \psi)^\ast$ for all vectors $\phi$ and $\psi\in \cH$;
\item the transition from the state  $|I\rangle$ in the state $\langle F |$
and the transition from the state $|\cT F \rangle$ to the state $\langle \cT I|$ 
are described with the same $S$-matrix:
\end{itemize}
\be
\cM_{I\ra F} =\langle F|S|I\rangle = \langle \cT I |S|\cT F\rangle = \cM_{\cT F\ra \cT I}.
\ee
Having in mind that
\be
\langle \cT I |S|\cT F\rangle =\ (\cT I,\; S\, \cT F)\ =\ (S^\dg\, \cT I, \;\cT F)\ 
=\ (\cT F,\; S^\dg\, \cT I)^\ast = \langle \cT F |S^\dg|\cT I\rangle^\ast,
\ee
we obtain the condition of $\cT$-invariance in terms of the decay and scattering amplitudes:
\be
\langle\ F |S|I \rangle \ =\ \langle\ \cT F |S^\dg |\cT I \rangle^\ast.
\ee
In perturbation theory, the $S$ matrix is expanded in the coupling constant $g$,
which is assumed to be a small parameter:
\be
S=1+ig T_1+g^2 T_2 +...
\ee
Here $T_1$ is nothing but the interaction Lagrangian, $T_1=\cL_{int}$; 
the unitarity of the $S$ matrix implies Hermiticity of the
interaction Lagrangian $T_1$. Thus the $\cT$-invariance condition
in the leading order of perturbation theory takes the form
\be
\ \langle\ \cT F |\cL_{int}|\cT I \rangle\ =\ \langle\ F |\cL_{int}|I \rangle^\ast.
\ee
To put it differently, if the complex-conjugated amplitude of the
transition between the states $I$ and $F$ differs {\bf in the leading order 
of perturbation theory} from the amplitude of the transition between 
the states $\cT I$ and $\cT F$ then the dynamics of such system 
is not invariant under the time reversal.

Let me illustrate this statement by considering the example of
the decay \kmung ; for definiteness, we consider the reference frame comoving
with the kaon. Let the average transverse\footnote{That is, transverse with respect to
the momenta of the outgoing particles} polarization of the muon $\xi\neq 0$.
This is possible only if the probabilities of the decay
into the states with positive and negative transverse polarizations of the muon 
differ from each other:
\be\label{eq:Probxipxim}
|\langle \mu(\vec k,\vec o) \nu(\vec k') \gamma(\vec q, \vec \epsilon)|S|K\rangle|^2
\neq |\langle \mu(\vec k,-\vec o) \nu(\vec k') \gamma(\vec q, \vec \epsilon)|S|K\rangle|^2.
\ee
Note that the state 
$ \langle\mu(\vec k,-\;\vec o) \nu(\vec k') \gamma(\vec q, \vec \epsilon)|$
can be obtained from the state
\[ \langle\mu(\vec -k,-\;\vec o) \nu(\vec -k') \gamma(\vec -q, \vec -\epsilon)|
= \langle\cT\;(\mu(\vec k,\vec o) \nu(\vec k') \gamma(\vec q, \vec \epsilon))| \]
as the result of the rotation by the angle of $180^o$ in the reaction plane.
For this reason,
\be\label{eq:TTrAmplMNG}
\langle \mu(\vec k,-\xi) \nu(\vec k') \gamma(\vec q, \vec \epsilon)|S|K\rangle =
\langle \cT\; \mu(\vec k,\xi) \nu(\vec k') \gamma(\vec q, \vec \epsilon)|S|\cT\,K\rangle.
\ee
The equations (\ref{eq:Probxipxim}) and (\ref{eq:TTrAmplMNG}) imply the
conclusion as follows: if the transverse polarization of the muon 
emerges in the first order of perturbation theory, then 
\be
\langle\cT\;\mu\nu\gamma|\cL_{int}|\cT\,K\rangle \neq \langle\mu\nu\gamma|\cL_{int}|K\rangle,
\ee
that is, the dynamics is not invariant under the time reversal.
 
However, we should take into account the following reasoning.
Since the transverse polarization of the muon is determined by the 
imaginary part of the decay amplitude, which does not vanish in higher
orders of perturbation theory due to unitarity condition,
the transverse polarization of the muon emerges in higher orders
even in the case of $\cT$-even interactions.

Thus the transverse polarization of the muon in the decay  \kmung \ 
can be caused by both electromagnetic and $\cT$-odd interactions
(beyond the SM)
\be
\xi = \xiem + \xio,
\ee

where $\xiem$ is the electromagnetic contribution to the transverse
polarization of the muon and $\xio$ is the contribution of 
the $\cT$-odd (and, therefore \CP-odd) interactions.

The \CP-violating interactions can be accounted for by the imaginary parts of
the coupling constants in the effective quark--lepton Lagrangian 
\bea\label{eq:EffQuarkLeptInt}
{\cal L}_{eff}&=&-{G_F \over \sqrt{2}}sin\theta_c \bar{s}\gamma^{\alpha}
(1-\gamma_5) u
\bar{\nu}\gamma_{\alpha}(1-\gamma_5)\mu+ \\ \nonumber 
&& +G_S \bar{s}u \bar{\nu}(1+\gamma_5)\mu + G_P \bar{s}\gamma_5 u \bar{\nu}(1+\gamma_5)\mu + \nonumber \\
&& + G_V \bar{s}\gamma^{\alpha}u \bar{\nu}\gamma_{\alpha}(1-\gamma_5)\mu
 +G_A \bar{s}\gamma^{\alpha}\gamma_5 u\bar{\nu}\gamma_{\alpha}(1-\gamma_5)\mu+\nonumber \\
&& +G_T \bar{s}\sigma^{\alpha\beta}(1-\gamma_5) u\bar{\nu}\sigma_{\alpha\beta}(1-\gamma_5)\mu \;+\; \mbox{H.c.}.  \nonumber 
\eea
The interactions in (\ref{eq:EffQuarkLeptInt}) arise from new physics.
Nonvanishing imaginary parts in the effective coupling constants
$G_P, G_V, G_A, G_T$  gives rise to the imaginary parts of the form
factors $F_{IB}, F_A, F_V, F_T$
parameteriznig the matrix element of the decay $K^+(p) \ra \mu^+(k)\nu(k')\gamma(q)$):

\newpage
\begin{figure*} \hbox{
\hspace*{-5pt}
       \epsfxsize=450pt \epsfbox{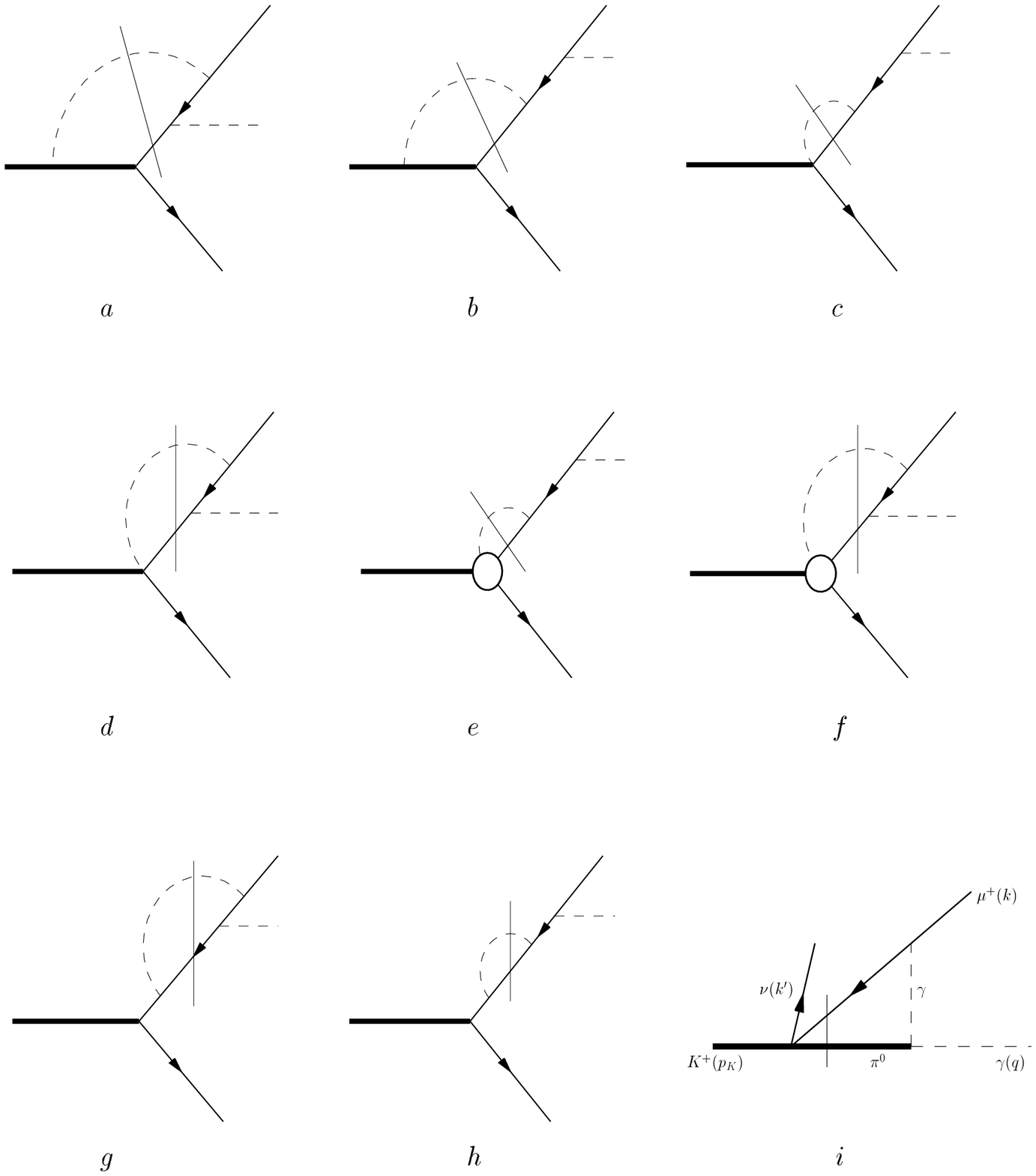} \hspace*{10pt}
       }
\caption{Diagrams giving a contribution to the imaginary part of the 
amplitude of the decay \kmung.} \label{fig:10}
\end{figure*}

Current limitations on the $\cT$-violation parameters in various extensions of the SM
allow the transverse polarization of the muon in the decay
\kmung \ to be rather large: the left-right symmetric models
based on the symmetry group $SU(2)_L\times SU(2)_R \times U(1)_{B-L}$
with one doublet $\Phi$ and two triplets $\Delta_{L,R}$ of Higgs bosons
can give $\xio \sim 7\times 10^{-3}$, 
supersymmetric models---$\xio \sim 5\times 10^{-3}$, 
leptoquark models---$\xio \sim 5 \times 10^{-3}$ \cite{Geng}.

The respective $\cT$-even contribution to the transverse polarization of the muon
is determined by the imaginary part of the decay amplitude (see Fig.~\ref{fig:10}) 
and emerges in the second order in $\alpha_{em}$.
Straightforward computations of $\xiem$ were made by several authors;
recent results \cite{Rogalyov} agree with each other and give the average value
$\langle \xiem \rangle \sim 0.5\times 10^{-3}$ (with the photon cutoff energy $\sim 25$~MeV).
The previous computations \cite{Kudenko} are incomplete:
either diagrams in Fig.~\ref{fig:10}$g$--\ref{fig:10}$h$ or the diagrams in Fig.~\ref{fig:10}$a$--\ref{fig:10}$h$
were not taken into account.

A similar $\cT$-even contribution to the correlation $\eta=\bf (\vec p_1, [\vec p_2 \times \vec p_3])$
in the decay $K^+ \ra \pi^0\mu_+\nu\gamma$ is given by similar diagrams and has 
the same order of magnitude $1.1\times 10^{-4}$ \cite{Braguta}.

\subsection{ \CP\ and $\cT$ violation in the decays $K\ra 3\pi$.}

The imaginary part of the effective coupling constants $g_8$, $g_{27}$ and $g_{ew}$
in the effective weak Lagrangian (\ref{eq:EffMesonLag})
gives rise to the \CP-violating effects in the decays $K\ra 3\pi$.

The kinematical variables used to describe the decay
$K(p) \ra \pi_1(p_1)\pi_2(p_2)\pi_3(p_3)$ are as follows:
\[
s_i=(p-p_i)^2, \ \ X={(s_1-s_2)^2\over m_\pi^2}, \ \ Y={(s_3-s_0)\over m_\pi^2},
\]
where "3" is the "odd" pion in either the $\tau (\pi^\pm\pi^\mp\pi^\mp)$ or 
$\tau' (\pi^\pm\pi^0\pi^0)$ decay mode. The slope parameters $g$ and $j$ 
are defined by the formula for the differential probability of the decay:
\be
|A(K\ra 3\pi)|^2 \sim 1+gY+jX+hY^2+kX^2.
\ee

The \CP-violating quantities are as follows:
\be
\delta_\Gamma={\Gamma(K^+ \ra 3\pi) - \Gamma(K^- \ra 3\pi) \over \Gamma(K^+ \ra 3\pi) + \Gamma(K^- \ra 3\pi)}.
\ee
and 
\be\label{eq:SlopeAsymm}
\delta_g={g(K^+ \ra 3\pi) - g(K^- \ra 3\pi) \over g(K^+ \ra 3\pi) + g(K^- \ra 3\pi)}
\ee
With the assumption that $\Delta I\leq 1/2$, the relevant $K\ra 3\pi$ amplitudes
can be expanded as follows:
\bea
A(K^+\ra \pi^+\pi^+\pi^-) &=& 2 a_c \left(1+i\alpha^0 +i{\alpha_0'\over 2}Y\right)
+[b_c(1+i\beta_0) +b_2(1+i\delta_0)]\,Y,   \\ \nonumber
A(K^+\ra \pi^0\pi^0\pi^+) &=& a_c \left(1+i\alpha^0 - i\alpha_0' Y\right)
-[b_c(1+i\beta_0) +b_2(1+i\delta_0)]\,Y.    \nonumber
\eea
From here on we restrict our attention to the slope asymmetry (\ref{eq:SlopeAsymm})
in the $\tau$ decay mode. As it usually is, this asymmetry is determined by the interplay
of {\it (i)} the imaginary parts of the parameters $a_c$ and $b_2$, stemming from
the \CP-odd effective weak Lagrangian (\ref{eq:EffMesonLag}) and {\it (ii)} the imaginary
part coming about the \CP-even final-state interactions \cite{Paver}:
\be
(\delta g)_\tau = {\alpha_0-\beta_0\over a_c(b_c+b_2)} \left( a_c\im b_c -b_c \im a_c) \right) 
 + {\alpha_0-\delta_0\over a_c(b_c+b_2)} \left( a_c\im b_2 -b_2 \im a_c) \right). 
\ee 
An evaluation of the strong rescattering phases in the one-loop approximation
of the $\chi$PT gives
\bea
\alpha_0&=& {\sqrt{1-(4m_\pi^2/s_0)} \over 32\pi F_\pi^2} (2 s_0 + m_\pi^2) \simeq 0.13, \\ \nonumber
\beta_0=-\delta_0&=& {\sqrt{1-(4m_\pi^2/s_0)} \over 32\pi F_\pi^2} ( s_0 - m_\pi^2) \simeq 0.05\, . \nonumber
\eea
The values $a_c$, $b_c$, $b_2$, $\alpha_0$, $\beta_0$, and $\delta_0$
can be expanded in powers of the $\chi$PT expansion parameter $\ds \lambda\simeq p/(4\pi F)$,
where $p$ defines the momentum scale and $F=93$~MeV. In the case of $K$-meson
decays, $\lambda\simeq 0.4$. In each order of the chiral expansion it is
helpful to isolate the  $\Delta I = 1/2$ and  $\Delta I = 3/2$ contributions to
$\Delta g$, the latter contribution being suppressed by the factor $\omega=0.045$.
The point is that, in the order $O(p^2)$, the neglect of the $\Delta I = 3/2$ contribution
gives rise to the relations $\ds a_c=-\, {1\over 3}\; {M_K^2\over M_\pi^2}\, b_c$
and $b_2=0$, which, in their turn imply that $\Delta g_{O(p^2),\; \Delta I = 1/2} =0.$
However, the $\Delta I = 1/2$ contribution does not vanish in the order $O(p^4)$
and so it dominates the total $O(p^4)$ contribution. It is natural to assume that
it is enhanced by the factor $\omega^{-1}=22.5$ (see \ref{eq:omega}) as compared to the  $\Delta I = 3/2$ contribution
in the order $O(p^4)$ of the $\chi$PT. The $O(p^4)$  $\Delta I = 3/2$ contribution is
suppressed by the $\chi$PT expansion parameter $\lambda^2$ as compared to
the $O(p^2)$  $\Delta I = 3/2$ contribution $A$. The above reasoning is
summarized in Table~5. \\[1mm]

{\large \bf Table 5.} Various contributions to $\ds \delta_g ={\Delta\!g\over 2g}$. \\[2mm]

\begin{tabular}{|c|c|c|c|}
\hline
&&& \\ 
Order of $\chi$PT & $\Delta I =1/2$ & $\Delta I =3/2$ & Numerical estimate \\ 
&&& \\ \hline

&&& \\ 
$O(p^2)$ & 0 & $A$ & $(1\div 3)\times 10^{-6}$ \\ 
&&& \\ \hline

&&& \\ 
$O(p^4)$ & $\ds \sim {\lambda^2 A\over \omega}$ & $ \sim \lambda^2 A$ &  $(0.4\div 1)\times 10^{-5}$\\ 
&&& \\ \hline

&&& \\ 
$O(p^6)$ & $\ds \sim {\lambda^4 A\over \omega}$ & $\sim \lambda^4 A $ & $ (0.8\div 2)\times 10^{-6}$ \\ 
&&& \\ \hline 

\end{tabular} 

\vspace*{2mm}
We see that the total contribution is dominated by the 
$O(p^4)$ contribution, which is $\omega^{-1}\lambda^2\simeq 4$ times greater than
the $O(p^2)$ contribution---due to vanishing of the $O(p^2)$  $\Delta I = 1/2$ contribution. 
Maiani and Paver \cite{Paver} assume that the enhancement factor may
run up to $10\div 20$. Therewith, the conclusion by Bel'kov {\it et al.}
\cite{Belkov} that the $O(p^6)$ corrections increase the enhancement
factor by the order of magnitude appears, in view of the above reasoning,
highly questionable. It should also be noticed that the multi-Higgs models
allow a two-fold increase of the parameter $\delta_g$ as compared with
the SM prediction \cite{Shabalin}.\\[1mm]

{\cha Acknowledgment:} I am grateful to G.G. Volkov for stimulating discussions.

\end{document}